\documentclass{article}
\usepackage[utf8]{inputenc}
\usepackage[letterpaper, margin=1in]{geometry}
\usepackage{graphicx}
\usepackage[section]{placeins}
\usepackage{subcaption}
\usepackage{float}
\usepackage{multicol}
\usepackage{multirow}
\usepackage[sorting=none]{biblatex}

\def\MC{MuCol}
\def\mm{$\mu^+\mu^-$}
\def\ten34{10$^{34}$ cm$^{-2}$ s$^{-1}$~}

\newcommand\snowmass{\begin{center}\rule[-0.2in]{\hsize}{0.01in}\\\rule{\hsize}{0.01in}\\
\vskip 0.1in Submitted to the  Proceedings of the US Community Study\\ 
on the Future of Particle Physics (Snowmass 2021)\\ 
\rule{\hsize}{0.01in}\\\rule[+0.2in]{\hsize}{0.01in} \end{center}}

\begin{document}

\begin{titlepage}

\snowmass

\begin{flushright}
March 2022
\end{flushright} 

\begin{center} 
{\Large 
Prospects for Heavy WIMP Dark Matter \\Searches at Muon Colliders
}
\end{center}

\begin{center} 
K. Black, T. Bose, Y. Chen, S. Dasu, H. Jia, 
\\D. Pinna, V. Sharma, N. Venkatasubramanian, C. Vuosalo
\\{\bf University of Wisconsin}
\end{center}

\begin{abstract}
%Prospects for Heavy WIMP Dark Matter Searches at Muon Colliders are great!
A future high-energy muon-muon collider could greatly extend our quest for new physics, providing cleaner final states than those produced at hadron colliders. Among its possible physics program, an interesting opportunity is provided by dark matter. Although strong astrophysical evidence indicates the existence of dark matter, there is no evidence yet for its non-gravitational interactions with standard model particles. If present, these interactions can be studied at colliders and in particular at the future muon-muon collider. 
In this whitepaper, we present a study for heavy weakly interacting massive particle dark matter particles that are part of a new electroweak multiplet and have a high mass. In particular,  we report on prospects for dark matter discovery both mono-photon and mono-Z processes.

\end{abstract}

\end{titlepage}

\section{Introduction}

A future high-energy Muon Collider (\MC)  could greatly extend our quest for new physics \cite{mucol}. At center of mass energies above 3 TeV and luminosities above \ten34, \mm$-$ collisions provide significantly cleaner final states than those produced at the HL-LHC. Even though the idea of a muon collider is not new, as it was first proposed in the late ‘60s~\cite{tikhonin1968effects, budker1982accelerators} and studied in detail by the Muon Accelerator Program (MAP), it is now receiving renewed interest because of its potential to overcome key limitations of other proposed collider concepts. 
%The enormous physics potential of colliding \mm~beams has sparked a wave of studies aimed at quantifying a possible physics program. 
For example, circular \mm~colliders could reach multi-TeV regime, within a limited spatial footprint and power budget because synchrotron radiation from muons is significantly lower than that of electrons. Furthermore, \mm~collisions are expected to take place with a small energy spread and lead to an improved energy resolution for physics measurements. 

The enormous physics potential of colliding \mm~beams has sparked a wave of studies aimed at quantifying a possible physics program~\cite{AlAli:2021let}. 
%There has been much attention paid to the physics prospects at \MC~recently~\cite{AlAli:2021let}. 
In this paper we explore the prospects for unearthing evidence for heavy weakly interacting massive particle (WIMP) dark matter (DM). 
%Dark Matter (DM) remains one of the outstanding mysteries of modern physics. 
Even though its existence is currently well established via various cosmological observations~\cite{Bertone:2004pz, Feng:2010gw, Porter:2011nv}, its nature is yet to be determined, making DM one of the outstanding mysteries of modern physics. According to the Standard Model (SM) of cosmology, in the total cosmic energy budget, our known matter only occupies about 4.9\%, the DM occupies 26.8\%, and the remaining is assumed to be dark energy. Although strong astrophysical evidence indicates the existence of DM, there is no evidence yet for non-gravitational interactions between DM and SM particles. This possible type of interactions have been studied in various type of experiments, such as direct detection~\cite{Cushman:2013zza} and indirect detection~\cite{Buckley:2013bha} experiments.
%Recent DM searches have exploited a number of direct detection~\cite{Cushman:2013zza} experiments and indirect detection~\cite{Buckley:2013bha}. 
The currently favoured possibility is that DM candidate may take a form of weakly interactive massive particles (WIMPs). Since there is no strong constraint on the WIMP mass, the searches for WIMP must continue beyond the HL-LHC reach, focusing on the high-mass region.

The WIMP model~\cite{Han:2020uak} considered in this paper includes a new electroweak (EW) multiplet with a high mass. The dark matter candidate is presumably the lowest mass state of this new multiplet and it can be either a neutral or a charged particle. The WIMP mass ($m_\chi$) set by the requirement of saturating the thermal relic abundance is in the range of 1-23 TeV. The model independent DM searches at the LHC are expected to saturate at few hundred GeV, even after the high-luminosity phase of the LHC program completes\cite{Low:2014,Han:2018,Vidal:2019}. 

In this study, we report on prospects for WIMP discovery at \mm~colliders with center of mass energies ($E_{CM}$) between 3 and 30 TeV and integrated luminosities ranging from 1 to 10 ab$^{-1}$. We have explored the possibility of discoveries in both mono-photon and mono-Z (both leptonic and hadronic decays), considering only the most significant backgrounds. Although there are several multiplets of the SM group with varying cross sections for production that could be considered, we limit ourselves to color-singlet-electroweak-doublet case in this initial study. 

\section{Simulation Setup and Analysis}

The signal is generated using the FeynRules~\cite{Alloul:2013bka} based model for Madgraph5~\cite{Alwall:2014hca} provided by the authors of Ref.~\cite{Han:2020uak}. Madgraph5, Pythia8~\cite{Bierlich:2022pfr} and Delphes~\cite{deFavereau:2013fsa} programs are setup for various \MC~configurations and used to produce mono-photon and mono-z signals for various dark matter particle masses and centers of mass energies of the collider, as well as the SM background processes. 
%We report on the significance of measurements of mono-photon and mono-Z event signatures of DM production over the Standard Model (SM) backgrounds. 

The Delphes program is setup with the Muon Collider data cards. The \MC-detector  acceptance effects and resolutions are used in selecting the final state particles, and smearing them appropriately. It is well-known that the beam induced background (BIB) due to downstream and upstream muon decays causes a very large deposition of particle shower energy in the detector. The BIB mitigation algorithms are currently in development. The BIB effects are completely ignored in the Delphes simulation that we have used. 

Since the dark matter WIMP exits the detector without interacting, the events are expected to have large missing transverse momentum (MET), 

\[
\mathrm{MET} = \sqrt{\Big[\sum_i{{p_x^i}\Big]^2 + \Big[\sum_i{p_y^i}\Big]^2}},
\]

and missing mass,

\[
m_{missing}^2 = {\Big(p_{\mu^+} + p_{\mu^-} - \sum_i{p_i}\Big)}^2,
\]

\noindent where $p_i$ are the four-vectors of the observed particles in the final states and $p_{\mu^+}$ and $p_{\mu^-}$ are the momenta of the colliding muons. Because any high-energy particles that exit the detector without interaction cause large MET and $m_{missing}$, we look for signal in that 2-dimensional plane. Because heavy WIMPs are produced in pairs and their masses considered in this analysis are high, the missing mass tends to have better sensitivity as the two WIMPs in the final state are close to back-to-back in the plane transverse to the beam direction.

The dominant backgrounds consist of events with $\gamma\nu\nu$ and Z$\nu\nu$ final states for mono-photon and mono-Z searches respectively. 
These processes can reproduce the typical final state expected from the signal. For the mono-photon topology the background processes can have a high energy photon from the initial state radiation. For the mono-Z case, the Z boson could be radiated from the final state neutrinos. The neutrinos are produced either by W-exchange between the muons or the radiation of a  Z-boson which decays to neutrinos. 
Further several subdominant processes in which some particles fall outside the acceptance of the detector (causing therefore a high MET in the event) are considered for the mono-Z case. There may be additional background from two-photon processes, which are neglected here. 

%The high energy photon is radiated from the initial state in case of the photon. The neutrinos are produced either by W-exchange between the muons or the radiation of a  $Z$-boson decays which decays to neutrinos. For the mono-$Z$ case, the Z boson could be radiated from the final state neutrinos as well. In both case, the neutrinos escape the detector resulting in mono-photon or mono-$Z$ events, identical to the signal events. Further several subdominant processes in which some particles fall outside the acceptance of the detector are considered for the mono-Z case. There may be additional background from two-photon processes, which are neglected here. 

The study is conducted for \MC~center-of-mass energies of 3, 6, 10 and 30 TeV with integrated luminosities of 1, 4, 10 and 10 ab$^{-1}$. 

\section{Mono-photon Analysis}

The mono-photon signal simulation consists of events dominated by \mm-annihilation diagrams with a Dirac SU(2) doublet and an accompanying high-energy photon emission from initial state radiation. The dominant background is from $\gamma\nu\nu$ production. 
Both types of processes were generated at various \mm-center-of-mass energies and signal events were generated considering different DM masses hypotheses and. The corresponding cross-sections are reported in Table~\ref{tab:mono-photon_xsec}.

%Both these types of processes were generated at various \mm-center-of-mass energies and for different DM masses for the case of signal events. The corresponding cross-sections are reported in Table~\ref{tab:mono-photon_xsec}.

\begin{figure}[H]
    \centering
    \begin{subfigure}{0.5\textwidth}
        \centering
        \includegraphics[width=3in]{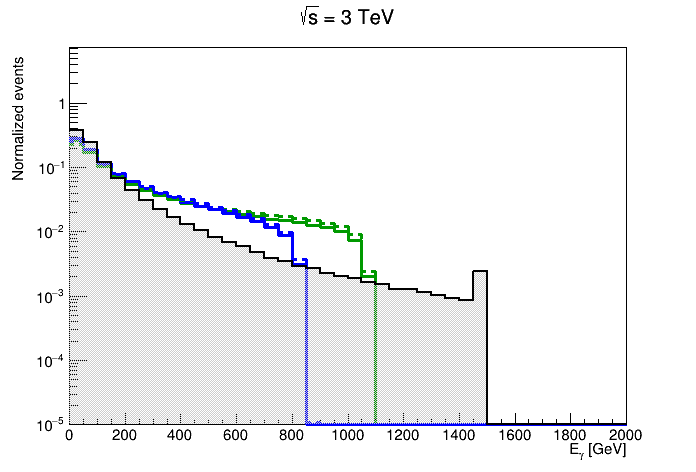}
        \caption{}
        \label{fig:monop_COM3_egamma}
    \end{subfigure}\hfill
    \begin{subfigure}{0.5\textwidth}
        \centering
        \includegraphics[width=3in]{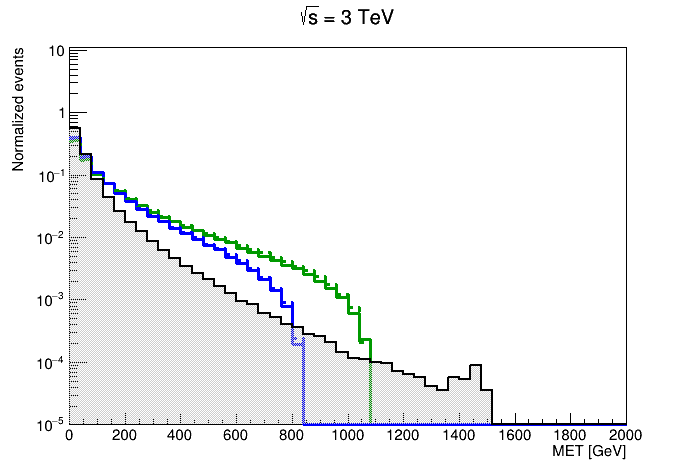}
        \caption{}
        \label{fig:monop_COM3_MET}
    \end{subfigure}\hfill
    \begin{subfigure}{0.5\textwidth}
        \centering
        \includegraphics[width=3in]{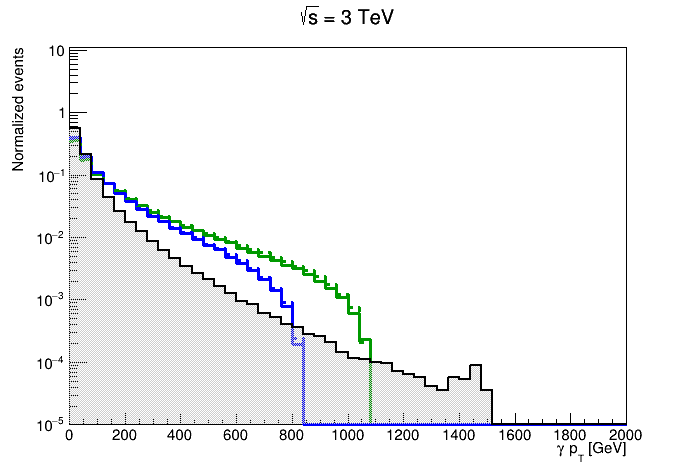}
        \caption{}
        \label{fig:monop_COM3_gammapT}
    \end{subfigure}\hfill
   \begin{subfigure}{0.5\textwidth}
        \centering
        \includegraphics[width=3in]{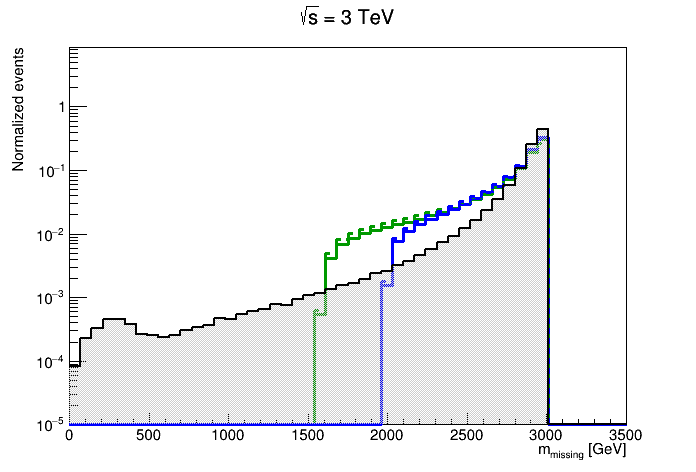}
        \caption{}
        \label{fig:monop_COM3_mm}
    \end{subfigure}\hfill
    \begin{subfigure}{0.5\textwidth}
        \centering
        \includegraphics[width=3in]{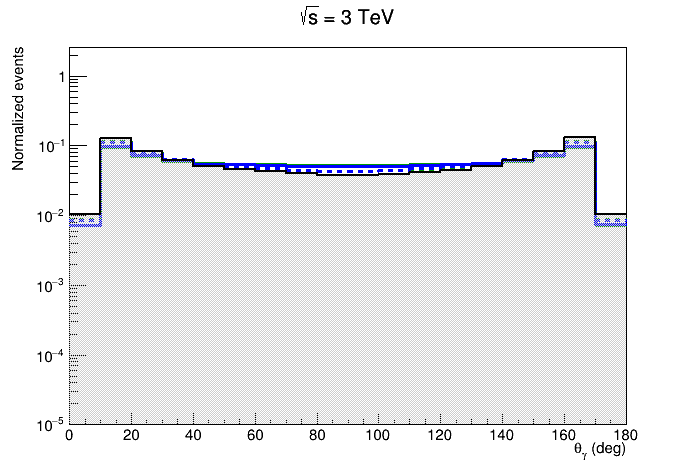}
        \caption{}
        \label{fig:monop_COM3_theta}
    \end{subfigure}\hfill
    \begin{subfigure}{0.5\textwidth}
        \centering
        \includegraphics[width=3in]{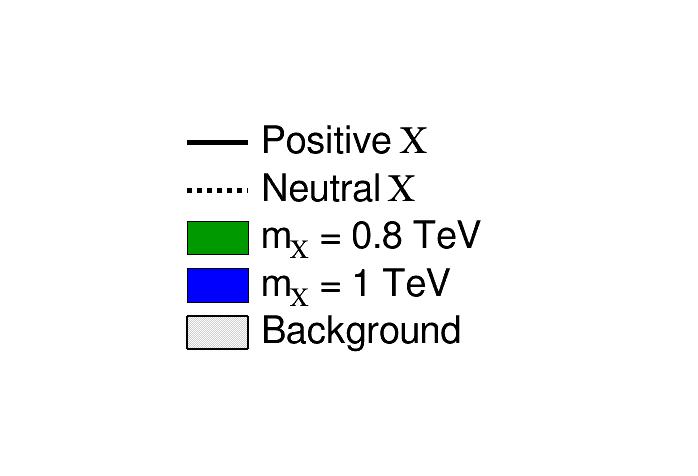}
        \label{fig:monop_COM3_legend}
    \end{subfigure}
    \caption{Normalized distributions for the photon energy E$_\gamma$ (a), MET (b), photon transverse momentum $\gamma \mathrm{p}_T$ (c), missing mass m$_{missing}$ (d), theta of the photon $\theta_{\gamma}$ (e) for different dark matter masses with both charged and neutral DM particles for a center of mass energy of 3 TeV after the requirement that at least one photon is present in the final state. %From the histograms, we see that applying the following selections will result in an improved signal sensitivity: $E_\gamma > 150\ \mathrm{GeV},\ MET > 75\ \mathrm{GeV},\ p_{T,\gamma} > 75\ \mathrm{GeV},\ 30^\circ <\theta_\gamma < 150^\circ$ 
    }
    \label{fig:monop-3}
\end{figure}

\begin{figure}[H]
    \centering
    \begin{subfigure}{0.5\textwidth}
        \centering
        \includegraphics[width=3in]{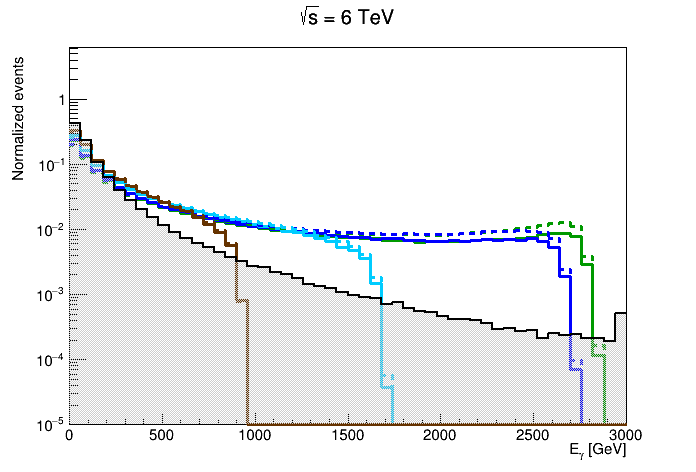}
        \caption{}
        \label{fig:monop_COM6_egamma}
    \end{subfigure}\hfill
    \begin{subfigure}{0.5\textwidth}
        \centering
        \includegraphics[width=3in]{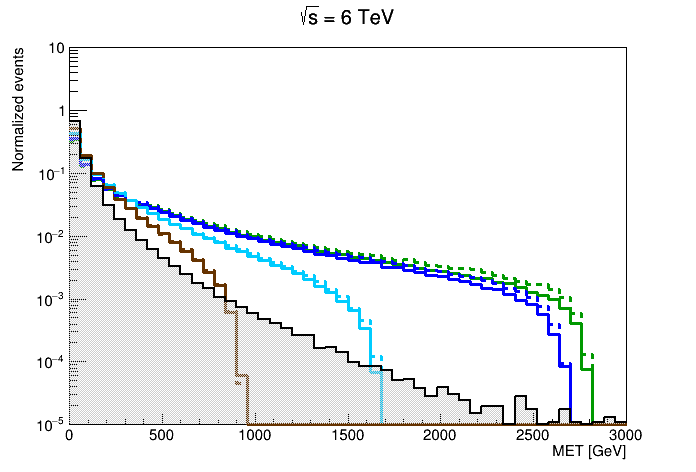}
        \caption{}
        \label{fig:monop_COM6_MET}
    \end{subfigure}\hfill
    \begin{subfigure}{0.5\textwidth}
        \centering
        \includegraphics[width=3in]{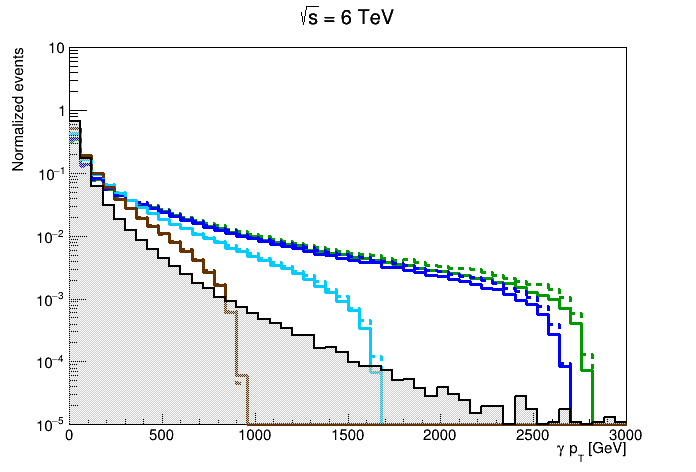}
        \caption{}
        \label{fig:monop_COM6_gammapT}
    \end{subfigure}\hfill
   \begin{subfigure}{0.5\textwidth}
        \centering
        \includegraphics[width=3in]{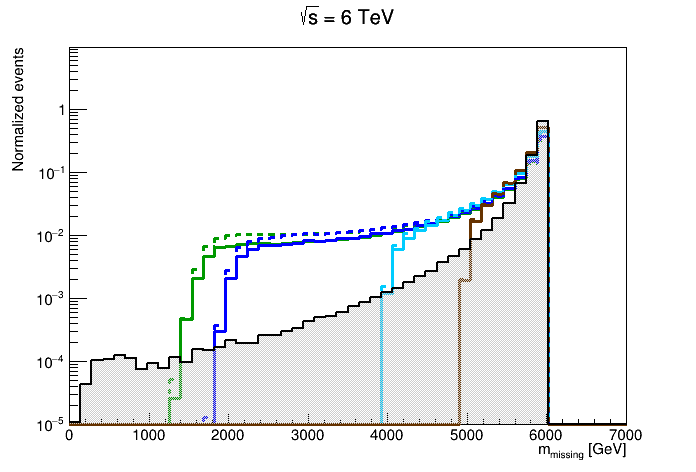}
        \caption{}
        \label{fig:monop_COM6_mm}
    \end{subfigure}\hfill
    \begin{subfigure}{0.5\textwidth}
        \centering
        \includegraphics[width=3in]{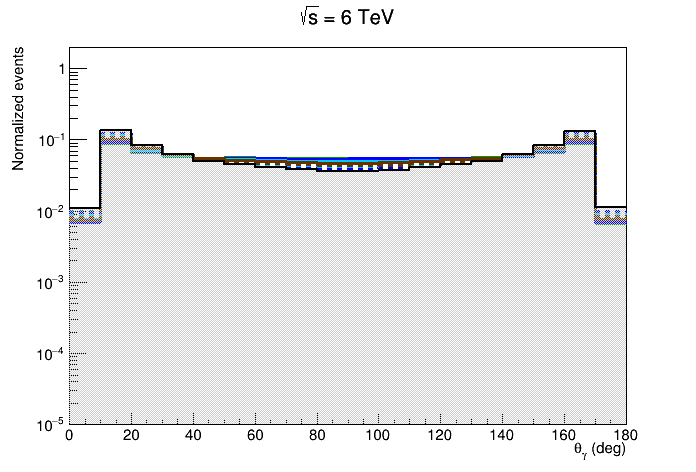}
        \caption{}
        \label{fig:monop_COM6_theta}
    \end{subfigure}\hfill
     \begin{subfigure}{0.5\textwidth}
        \centering
        \includegraphics[width=3in]{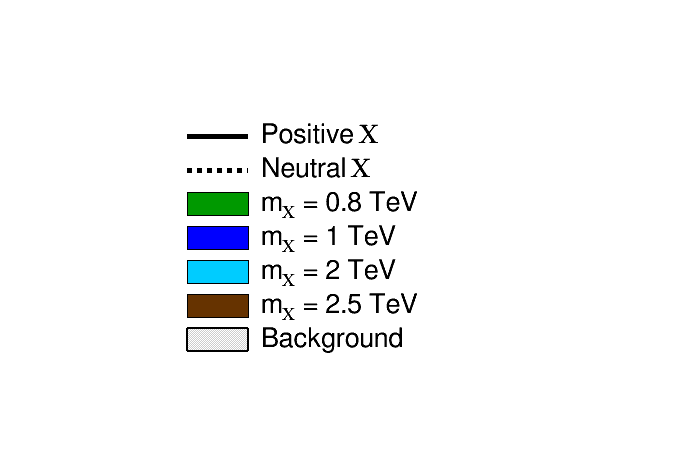}
        \label{fig:monop_COM6_legend}
    \end{subfigure}
    \caption{Normalized distributions for the photon energy E$_\gamma$ (a), MET (b), photon transverse momentum $\gamma \mathrm{p}_T$ (c), missing mass m$_{missing}$ (d), theta of the photon $\theta_{\gamma}$ (e) for different dark matter masses with both charged and neutral DM particles for a center of mass energy of 6 TeV after the requirement that at least one photon is present in the final state. %From the histograms, we see that applying the following selections will result in an improved signal sensitivity: $E_\gamma > 200\ \mathrm{GeV},\ MET > 100\ \mathrm{GeV},\ p_{T,\gamma} > 100\ \mathrm{GeV},\ 40^\circ <\theta_\gamma < 140^\circ$ 
    }
    \label{fig:monop-6}
\end{figure}

\begin{figure}[H]
    \centering
    \begin{subfigure}{0.5\textwidth}
        \centering
        \includegraphics[width=3in]{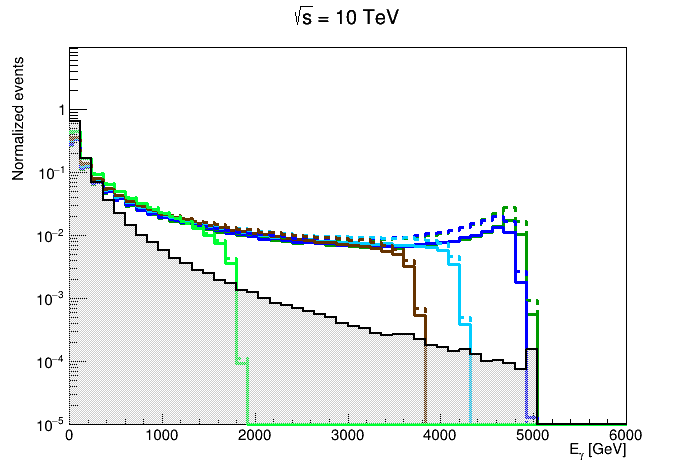}
        \caption{}
        \label{fig:monop_COM10_egamma}
    \end{subfigure}\hfill
    \begin{subfigure}{0.5\textwidth}
        \centering
        \includegraphics[width=3in]{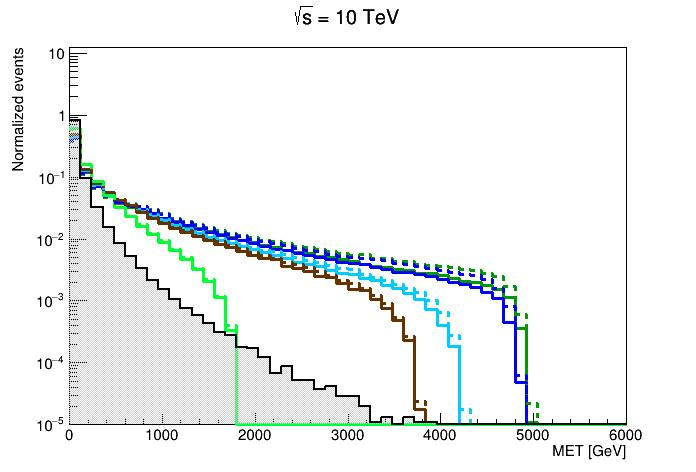}
        \caption{}
        \label{fig:monop_COM10_MET}
    \end{subfigure}\hfill
    \begin{subfigure}{0.5\textwidth}
        \centering
        \includegraphics[width=3in]{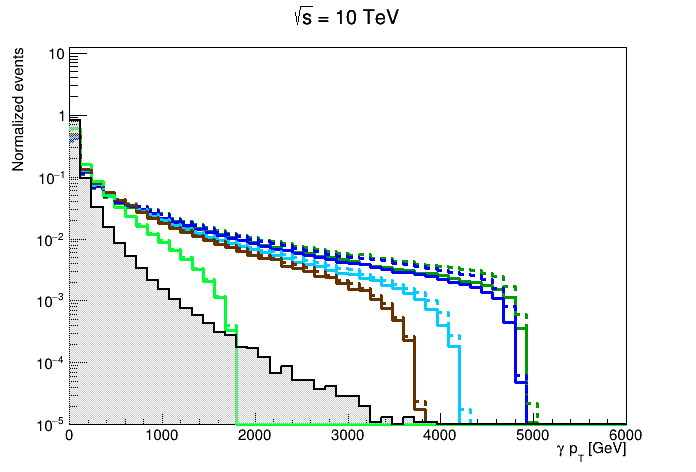}
        \caption{}
        \label{fig:monop_COM10_gammapT}
    \end{subfigure}\hfill
   \begin{subfigure}{0.5\textwidth}
        \centering
        \includegraphics[width=3in]{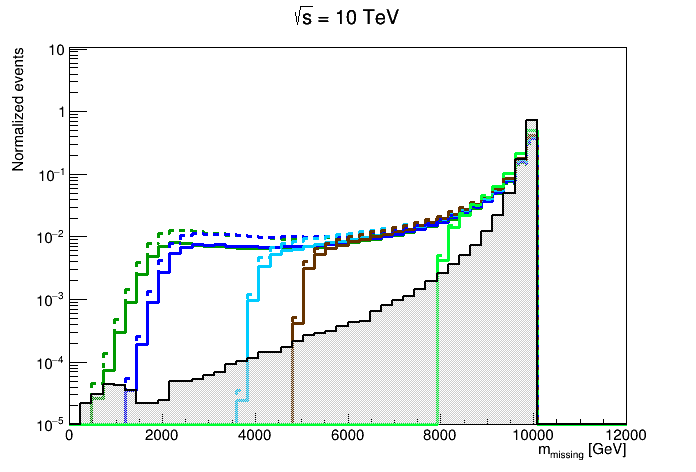}
        \caption{}
        \label{fig:monop_COM10_mm}
    \end{subfigure}\hfill
    \begin{subfigure}{0.5\textwidth}
        \centering
        \includegraphics[width=3in]{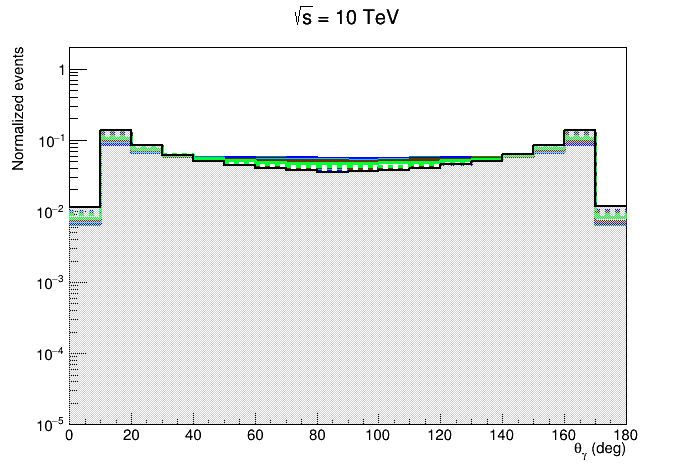}
        \caption{}
        \label{fig:monop_COM10_theta}
    \end{subfigure}\hfill
     \begin{subfigure}{0.5\textwidth}
        \centering
        \includegraphics[width=3in]{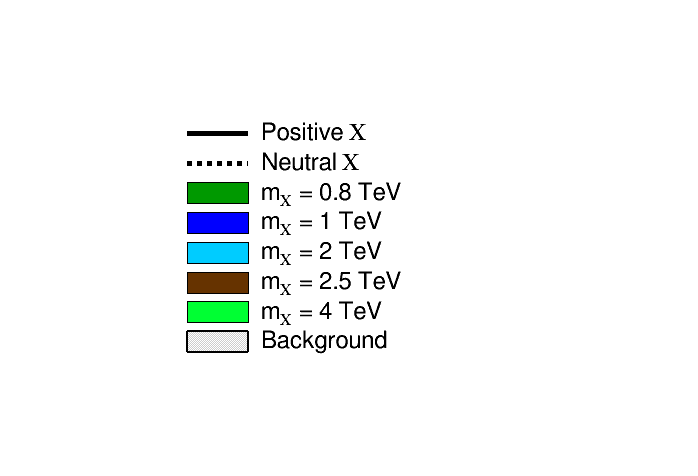}
        \label{fig:monop_COM10_legend}
    \end{subfigure}
    \caption{Normalized distributions for the photon energy E$_\gamma$ (a), MET (b), photon transverse momentum $\gamma \mathrm{p}_T$ (c), missing mass m$_{missing}$ (d), theta of the photon $\theta_{\gamma}$ (e) for different dark matter masses with both charged and neutral DM particles for a center of mass energy of 10 TeV after the requirement that at least one photon is present in the final state.% From the histograms, we see that applying the following selections will result in an improved signal sensitivity: $E_\gamma > 200\ \mathrm{GeV},\ MET > 100\ \mathrm{GeV},\ p_{T,\gamma} > 100\ \mathrm{GeV},\ 40^\circ <\theta_\gamma < 140^\circ$ 
    }
    \label{fig:monop-10}
\end{figure}

\begin{figure}[H]
    \centering
    \begin{subfigure}{0.5\textwidth}
        \centering
        \includegraphics[width=3in]{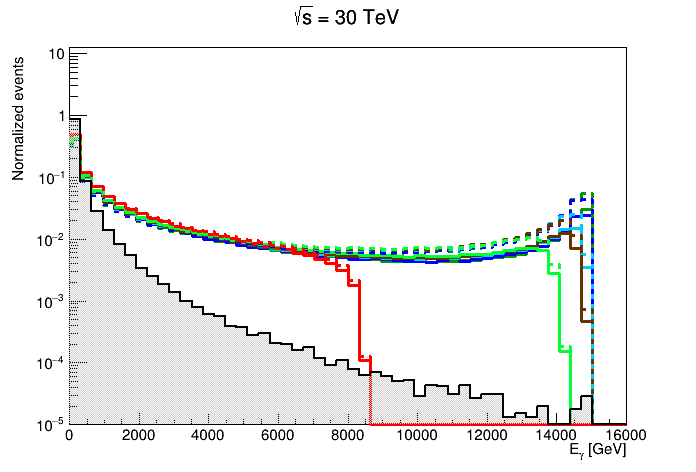}
        \caption{}
        \label{fig:monop_COM30_egamma}
    \end{subfigure}\hfill
    \begin{subfigure}{0.5\textwidth}
        \centering
        \includegraphics[width=3in]{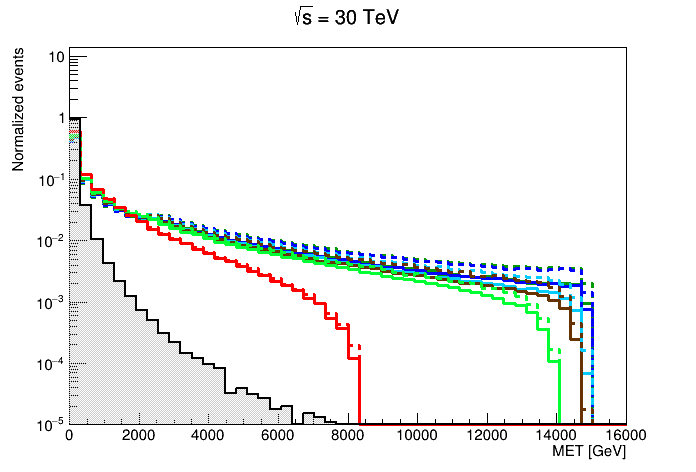}
        \caption{}
        \label{fig:monop_COM30_MET}
    \end{subfigure}\hfill
    \begin{subfigure}{0.5\textwidth}
        \centering
        \includegraphics[width=3in]{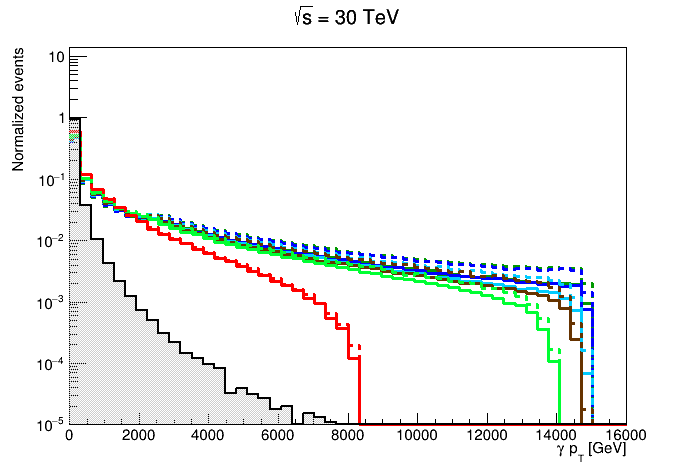}
        \caption{}
        \label{fig:monop_COM30_gammapT}
    \end{subfigure}\hfill
   \begin{subfigure}{0.5\textwidth}
        \centering
        \includegraphics[width=3in]{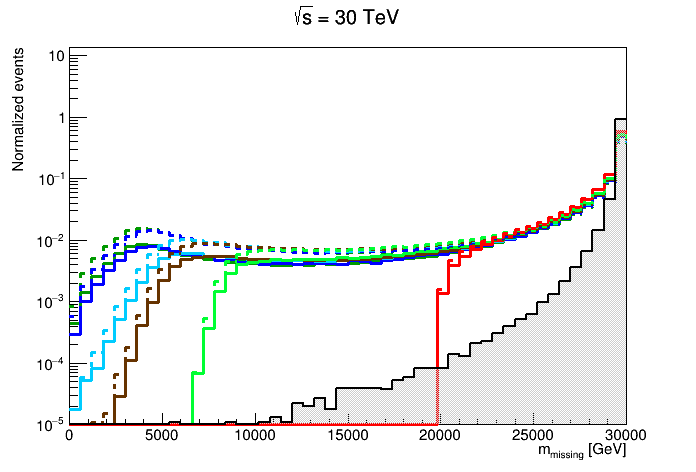}
        \caption{}
        \label{fig:monop_COM30_mm}
    \end{subfigure}\hfill
    \begin{subfigure}{0.5\textwidth}
        \centering
        \includegraphics[width=3in]{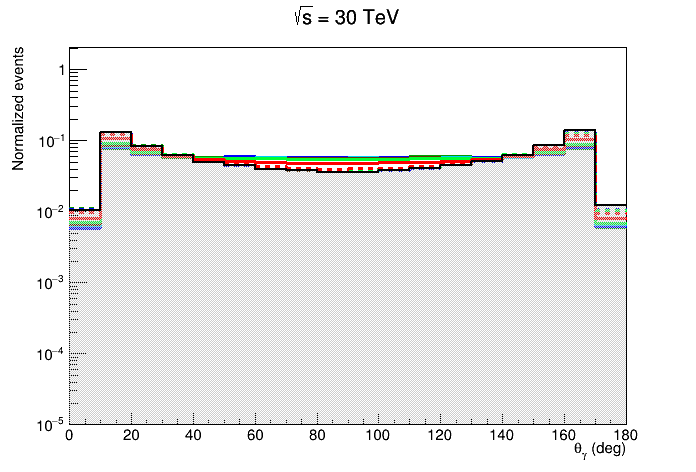}
        \caption{}
        \label{fig:monop_COM30_theta}
    \end{subfigure}\hfill
     \begin{subfigure}{0.5\textwidth}
        \centering
        \includegraphics[width=3in]{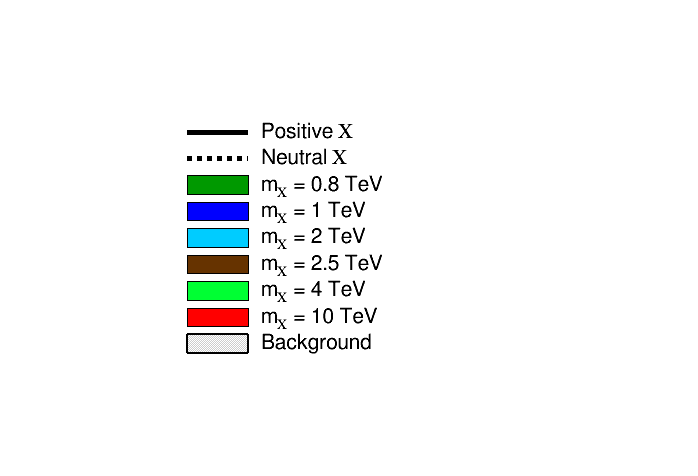}
        \label{fig:monop_COM30_legend}
    \end{subfigure}
    \caption{Normalized distributions for the photon energy E$_\gamma$ (a), MET (b), photon transverse momentum $\gamma \mathrm{p}_T$ (c), missing mass m$_{missing}$ (d), theta of the photon $\theta_{\gamma}$ (e) for different dark matter masses with both charged and neutral DM particles for a center of mass energy of 30 TeV after the requirement that at least one photon is present in the final state.% From the histograms, we see that applying the following selections will result in an improved signal sensitivity: $E_\gamma > 500\ \mathrm{GeV},\ MET > 500\ \mathrm{GeV},\ p_{T,\gamma} > 500\ \mathrm{GeV},\ 30^\circ <\theta_\gamma < 150^\circ$ 
    }
    \label{fig:monop-30}
\end{figure}

Simulated events are required to have at least a high energy photon in the final state. 
The following studies are based on what presented in Ref.~\cite{Han:2020uak}, with the extension of considering the full Delphes simulation for both signal and background processes to include the \MC-detector acceptance effects and resolutions, as well as an optimized variable selection to further increase the signal sensitivity of the search. 
Based on the studies~\cite{Han:2020uak}, different kinematic distributions were investigated to identify discriminating variables that can be employed to identify a phase space where signal events are enhanced with respect to background processes. 
The variables that were found to perform best for this discrimination are the following: energy of the photon (E$_\gamma$), MET, transverse momentum of the photon ($\gamma\mathrm{p}_{T}$), missing mass (m$_{missing}$) and the $\theta$ angle of the photon ($\theta_{\gamma}$). The normalized distributions for these variables are presented for different centers of mass energies of the collider and dark matter particle masses in Figures~\ref{fig:monop-3}-\ref{fig:monop-30}.

From these distributions, it can be inferred an event selection to be applied to improve the signal sensitivity of the mono-photon analysis. The final selection applied on the discriminating variables listed above is summarized in Table~\ref{tab:mono-photon_selections} for the different centers of mass energies considered for the collider.

The sensitivity of this analysis to mono-photon events is quantified in terms of a figure of merit (FOM) defined as: 

\begin{center}
    $FOM = \frac{\mathrm{s}}{\sqrt{\mathrm{b}}}$
\end{center}

where s is the numbers of expected signal events and b is the number of expected SM background events after the final selection. The FOM obtained for different centers of mass energies of the collider and dark matter masses are listed in Table~\ref{tab:mono-photon_FOM}.

\begin{table}[H]
    \centering
    \renewcommand{\arraystretch}{1.2}
    \begin{tabular}{|l|c|c|c|c|c|c|c|c|}
        \hline
        \multirow{2}{2.5cm}{$m_\chi$ / $\sqrt{s}~(\int{d{\cal L}})$}& \multicolumn{2}{c}{3 TeV (1 ab$^{-1}$)} & \multicolumn{2}{|c}{6 TeV (4 ab$^{-1}$)} & \multicolumn{2}{|c}{10 TeV (10 ab$^{-1}$)} & \multicolumn{2}{|c|}{30 TeV (10 ab$^{-1}$)}\\
        \cline{2-9}
        & $\sigma_{sig}$(fb) & $\sigma_{bkg}$(fb) & $\sigma_{sig}$(fb) &$\sigma_{bkg}$(fb) & $\sigma_{sig}$(fb) & $\sigma_{bkg}$(fb) & $\sigma_{sig}$(fb) & $\sigma_{bkg}$(fb)\\
        
        \hline
       &&&&&&&&\\
        0.8 TeV & 1.554 (c) & 2979 & 0.6492 (c) & 3160 & 0.3018 (c) & 3242 & 0.0511 (c) & 3304\\
       & 0.514 (n) && 0.1830 (n) && 0.0774 (n)&& 0.0111 (n)&\\
       &&&&&&&&\\
        1.0 TeV & 1.219 (c) & 2979 & 0.5952 (c) & 3160 & 0.2826 (c) & 3242 & 0.0488 (c) & 3304\\
        & 0.430 (n) && 0.1752 (n)&& 0.0749 (n) && 0.0109 (n) &\\
        &&&&&&&&\\
            2.0 TeV & n/a & n/a & 0.3655 (c) & 3160 & 0.2176 (c) & 3242 & 0.0415 (c) & 3304\\
        &&& 0.1285 (n)&& 0.0662 (n) && 0.0101 (n) &\\
        &&&&&&&&\\
        2.5 TeV & n/a & n/a & 0.223 (c) & 3160 & 0.1916 (c) & 3242 & 0.0393 (c) & 3304\\
        &&& 0.085 (n) && 0.0618 (n) && 0.0099 (n)&\\
        &&&&&&&&\\
        4.0 TeV & n/a & n/a & n/a & n/a & 0.1050 (c) & 3242 & 0.0341 (c) & 3304\\
        &&&&& 0.0393 (n) && 0.0094 (n) &\\
        &&&&&&&&\\
        10.0 TeV & n/a & n/a & n/a & n/a & n/a & n/a & 0.0202 (c) & 3304\\
        &&&&&&& 0.0071 (n) &\\
        \hline
    \end{tabular}
    \caption{Mono-photon signal ($\sigma_{sig}$) and background ($\sigma_{bkg}$) cross-sections (in fb) for muon colliders operating at various center of mass energies and integrated luminosity. Cross sections for charged DM particles are indicated with (c) and cross sections for neutral DM particles are indicated with (n). These cross section values were obtained after running the signal and background processes with 500,000 events each through MadGraph.}
    \label{tab:mono-photon_xsec}
\end{table}

From this study, we find that the mono-photon channel would be sensitive to color-singlet-electroweak-doublet DM candidates and have a significant impact in the search for the thermal dark matter. In order to target lower DM masses up to about $2$~TeV, muon colliders operating at low center of mass energies will be a preferred option, while to discover or constrain DM candidates above $2$~TeV higher energies and higher luminosities will be needed.

\begin{table}[H]
    \centering
    \begin{tabular}{|l|c|c|c|c|}
        \hline
        &&&&\\
        $\sqrt{s}~(\int{d{\cal L}})$ / Discrminating Variables & $\theta_{\gamma}$& $E_\gamma$ & MET & $\gamma_{p_T}$\\ 
        &&&&\\\hline
        &&&&\\
        3 TeV (1 ab$^{-1}$) & $>30^\circ$, $<150^\circ$ & $>150\ \mathrm{GeV}$ & $>75\ \mathrm{GeV}$ & $>75\ \mathrm{GeV}$\\
        &&&&\\
        6 TeV (4 ab$^{-1}$)  & $>40^\circ$, $<140^\circ$ & $>200\ \mathrm{GeV}$ & $>100\ \mathrm{GeV}$ & $>100\ \mathrm{GeV}$\\
        &&&&\\
        10 TeV (10 ab$^{-1}$) & $>40^\circ$, $<140^\circ$ & $>200\ \mathrm{GeV}$ & $>100\ \mathrm{GeV}$ & $>100\ \mathrm{GeV}$\\
        &&&&\\
        30 TeV (10 ab$^{-1}$) & $>40^\circ$, $<140^\circ$ & $>500\ \mathrm{GeV}$ & $>500\ \mathrm{GeV}$ & $>500\ \mathrm{GeV}$\\
        &&&&\\\hline
    \end{tabular}
    \caption{Final selections on the discriminating variables for various center of mass energies. Note that the missing mass selection $m_{missing}^2 > 4m_{\chi}^2$ is also applied to each center of mass energy. FOM calculations (see Table~\ref{tab:mono-photon_FOM}) are performed after applying these selections.}
    \label{tab:mono-photon_selections}
\end{table}

\begin{table}[H]
    \centering
    \begin{tabular}{|l|c|c|c|c|}
        \hline
        &&&&\\
        $m_\chi$ / $\sqrt{s}~(\int{d{\cal L}})$ & 3 TeV (1 ab$^{-1}$) & 6 TeV (4 ab$^{-1}$) & 10 TeV (10 ab$^{-1}$) & 30 TeV (10 ab$^{-1}$) \\ 
        &&&&\\\hline
        &&&&\\
        0.8 TeV & 1.02 & 1.01 & 0.81 & 0.23\\
        &&&&\\
        1.0 TeV  & 0.72 & 0.91 & 0.75 & 0.22\\
        &&&&\\
        2.0 TeV & n/a & 0.47 & 0.55 & 0.18\\
        &&&&\\
        2.5 TeV & n/a & 0.22 & 0.47 & 0.17\\
        &&&&\\
        4.0 TeV & n/a & n/a & 0.21 & 0.15\\
        &&&&\\
        10.0 TeV & n/a & n/a & n/a & 0.07\\
        &&&&\\\hline
    \end{tabular}
    \caption{Significance figure of merit for mono-photon signals as a function of dark matter mass ($m_\chi$) taking into consideration both neutral and charged DM particles, for muon colliders operating at various centers of mass and integrated luminosity.}
    \label{tab:mono-photon_FOM}
\end{table}

\section{Mono-Z Analysis}
\begin{figure}[htb]
    \centering
    \includegraphics[width=2.5in]{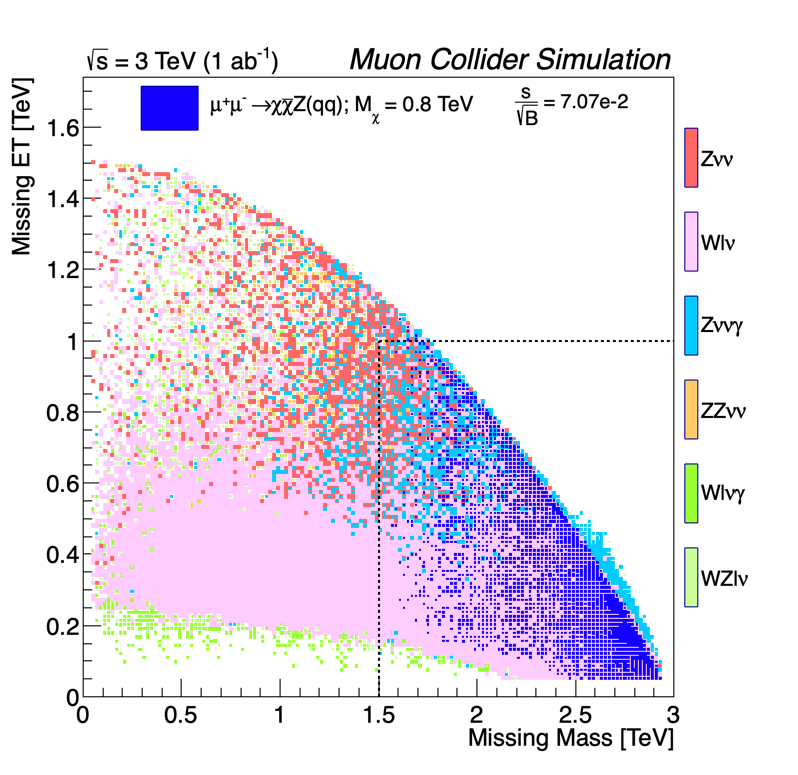}
    \includegraphics[width=2.5in]{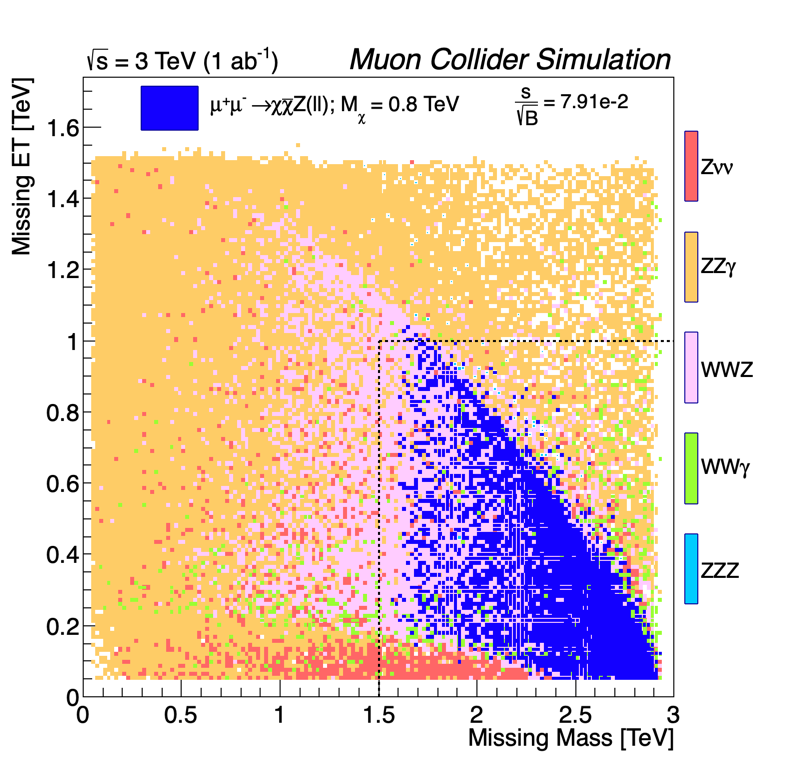}\\ 
    \includegraphics[width=2.5in]{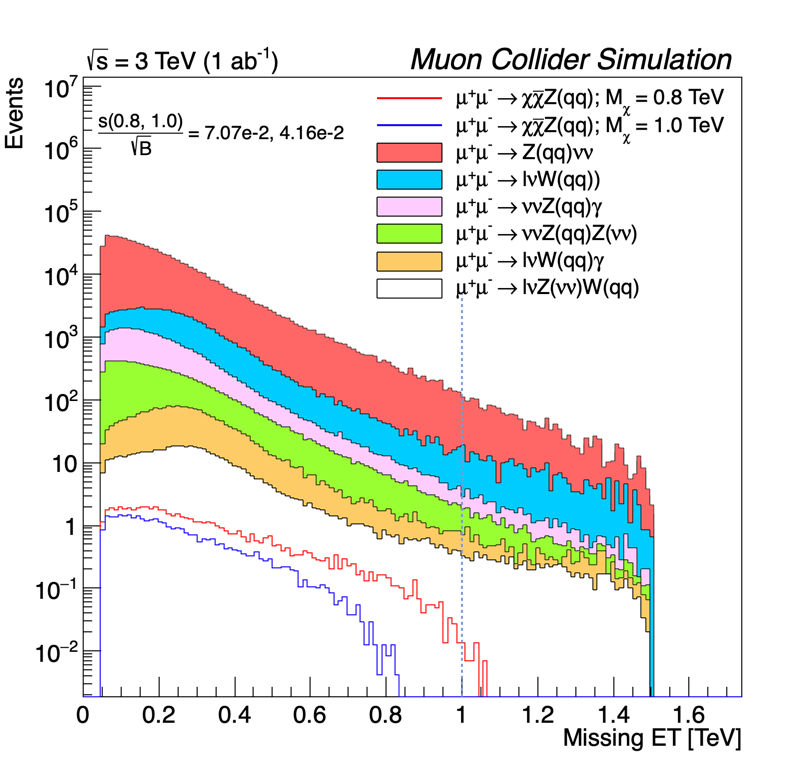} 
    \includegraphics[width=2.5in]{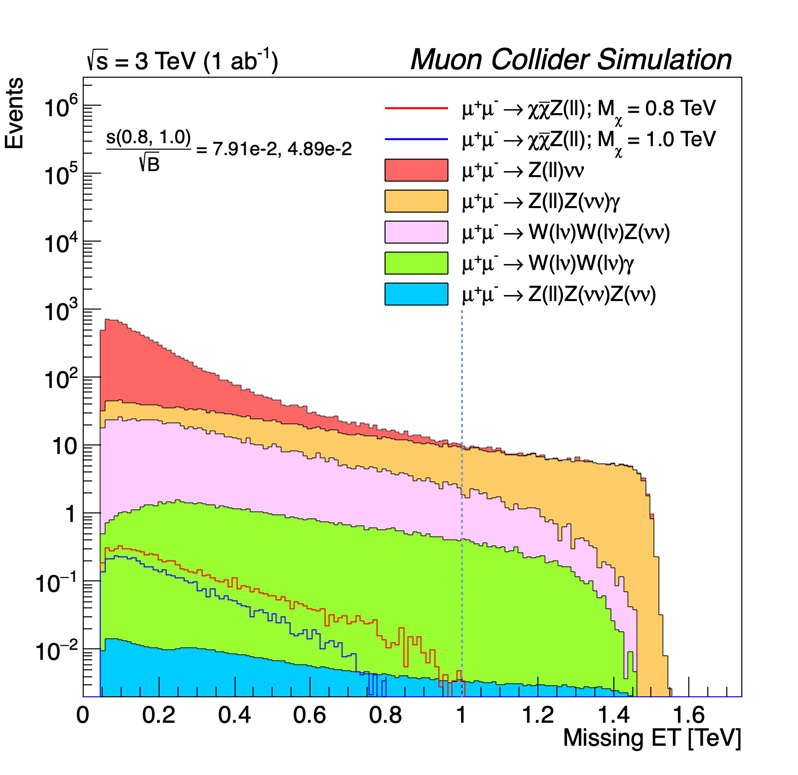} \\
    \includegraphics[width=2.5in]{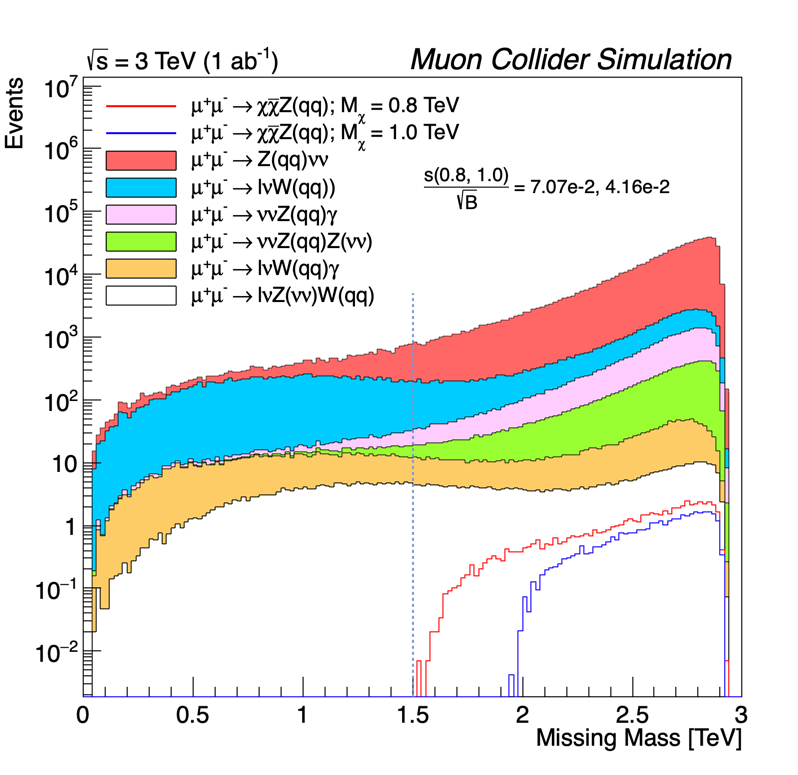}
    \includegraphics[width=2.5in]{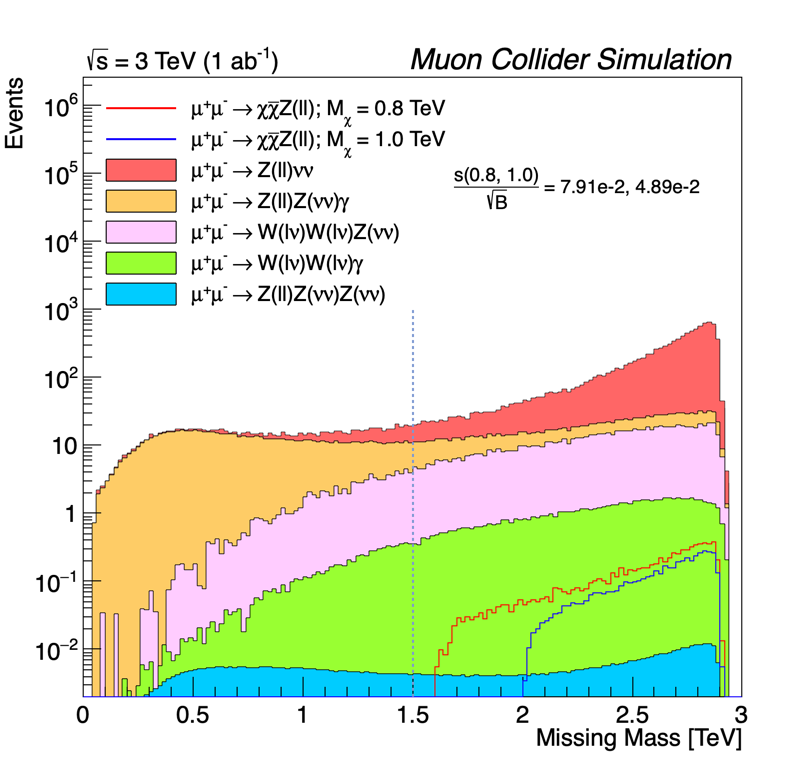}
    \caption{The left side of the figure has plots for mono-Z($qq$) while right side has for mono-Z($\ell\ell$). Top panel: Scatter plot in missing transverse momentum versus missing mass plane for $m_\chi=0.8$ GeV, with proposed cuts indicated as the black rectangle. The simulated signal events in both scatter plot are shown in blue color and various simulated backgrounds are indicated in other colors. The signal significance (s/$\sqrt{B}$ is shown for an integrated luminosity of $1$ ab$^{-1}$ for the simulated $\sqrt{s}=3$ TeV data. The middle and bottom panel shows the MET and the missing mass projections. The signal and backgrounds in the MET and missing mass distributions are normalized to respective cross section and chosen value of luminosity.}
    \label{fig:monoz-s3}
\end{figure}

\begin{figure}[htb]
    \centering
    \includegraphics[width=2.5in]{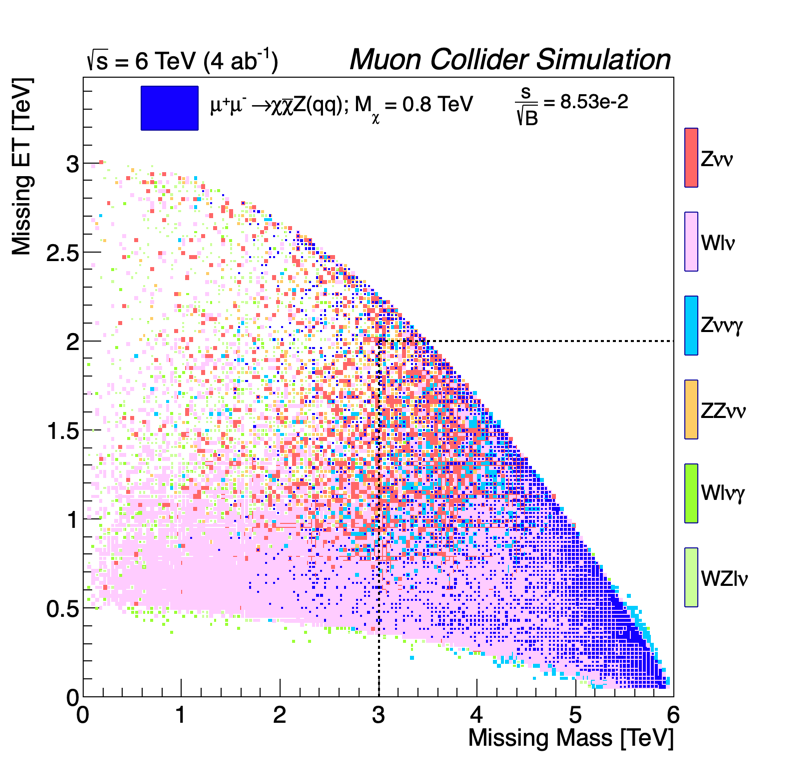}
    \includegraphics[width=2.5in]{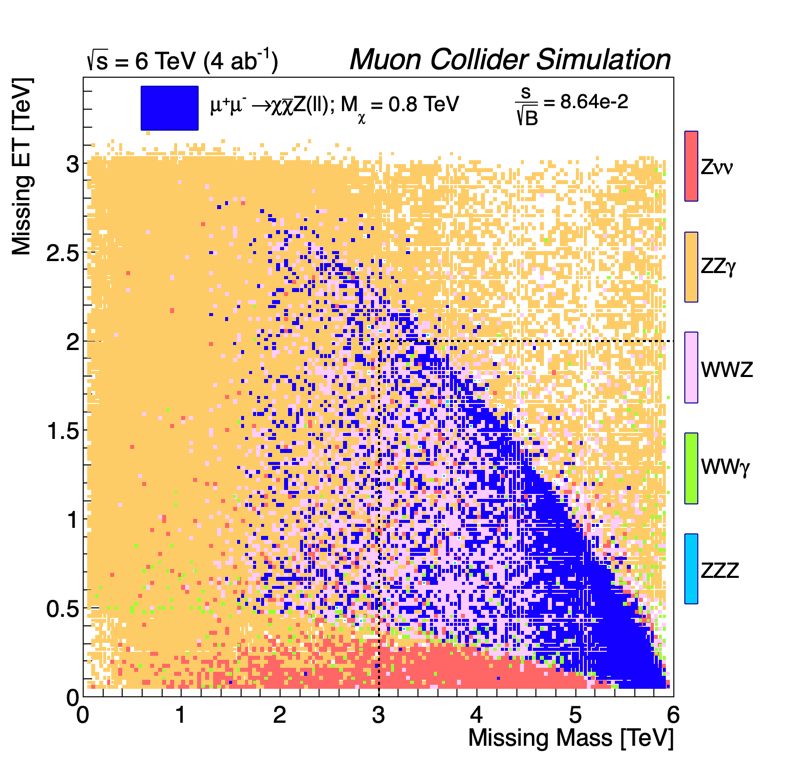}\\
   \includegraphics[width=2.5in]{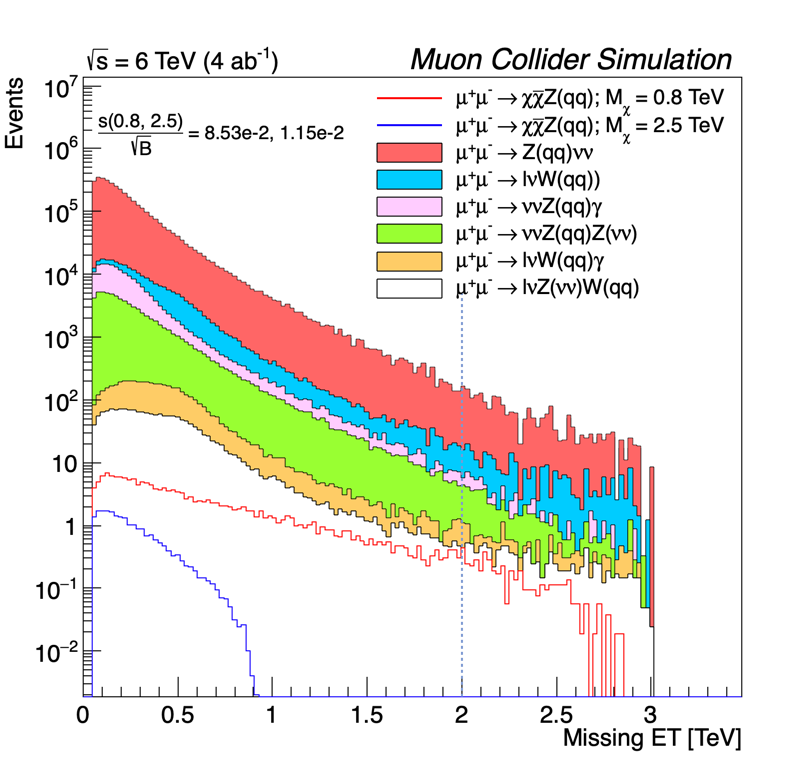}
    \includegraphics[width=2.5in]{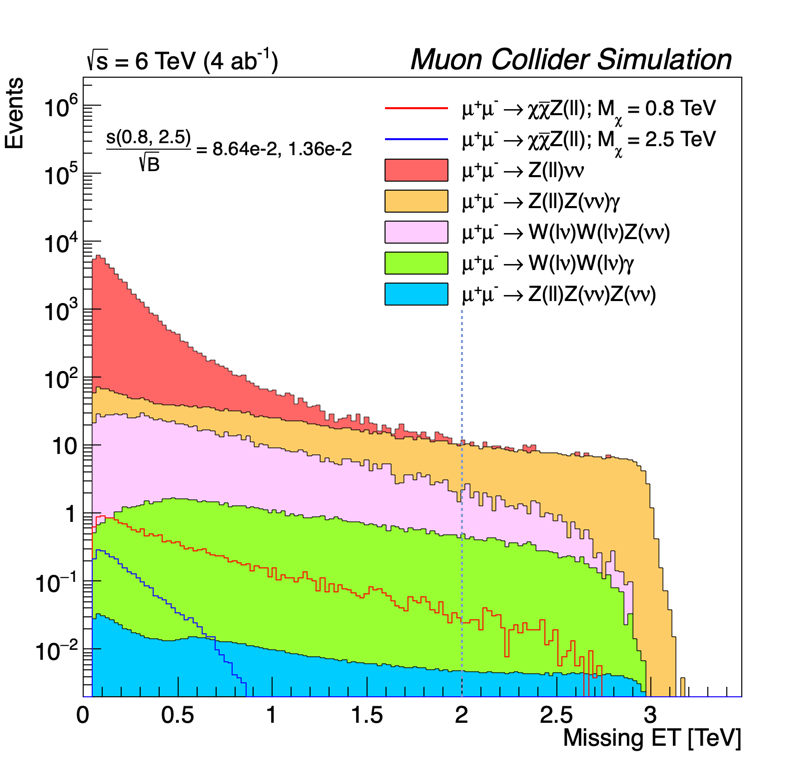}\\
    \includegraphics[width=2.5in]{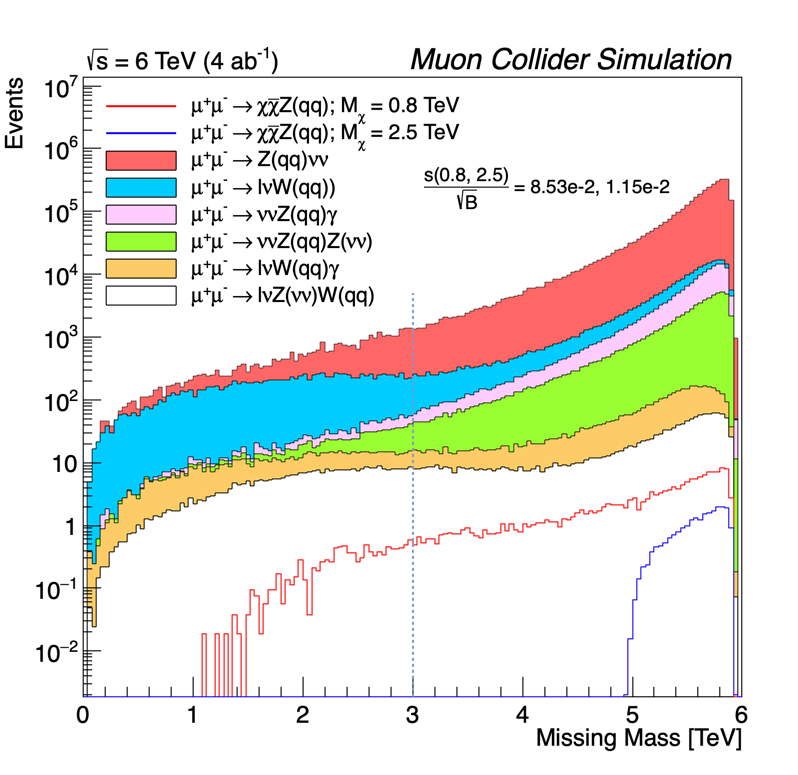}
    \includegraphics[width=2.5in]{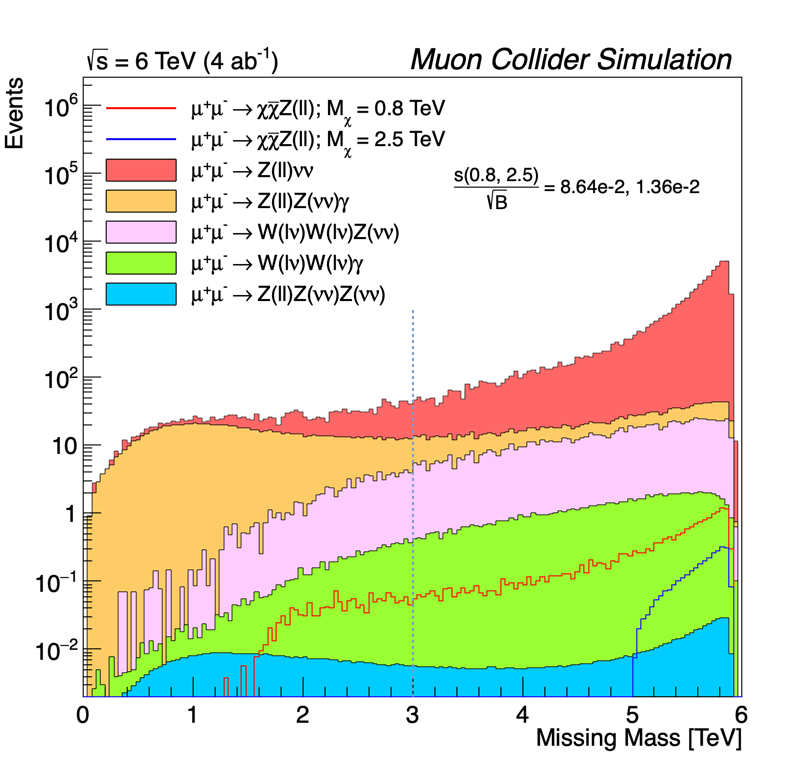}
    \caption{The left side of the figure has plots for mono-Z($qq$) while right side has for mono-Z($\ell\ell$). Top panel: Scatter plot in missing transverse momentum versus missing mass plane for $m_\chi=0.8$ GeV, with proposed cuts indicated as the black rectangle. The simulated signal events in both scatter plot are shown in blue color and various simulated backgrounds are indicated in other colors. The signal significance (s/$\sqrt{B}$ is shown for an integrated luminosity of $4$ ab$^{-1}$ for the simulated $\sqrt{s}=10$ TeV data. The middle and bottom panel shows the MET and the missing mass projections. The signal and backgrounds in the MET and missing mass distributions are normalized to respective cross section and chosen value of luminosity}
    \label{fig:monoz-s6}
\end{figure}

\begin{figure}[htb]
    \centering
    \includegraphics[width=2.5in]{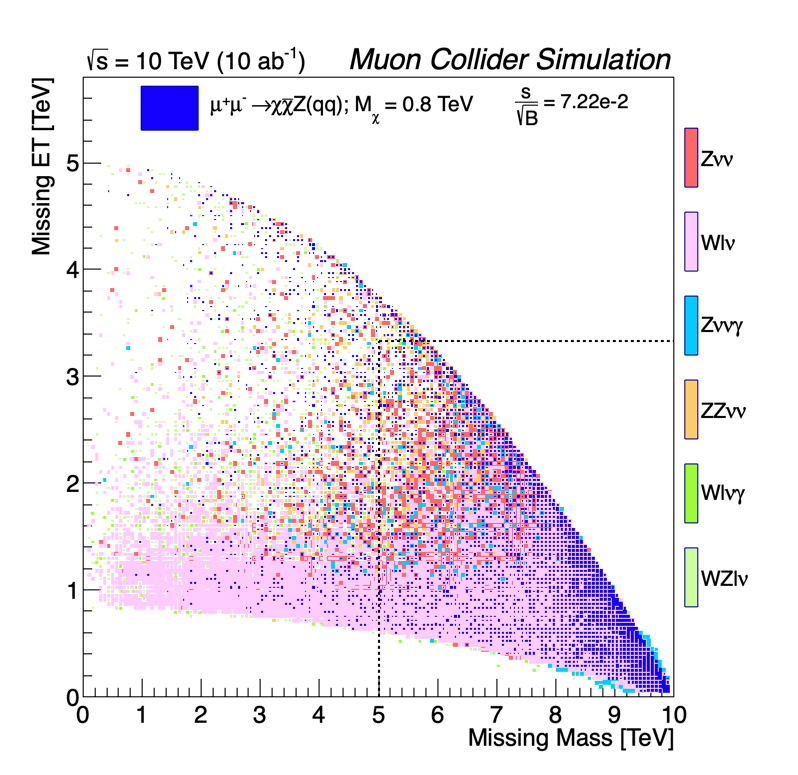}
    \includegraphics[width=2.5in]{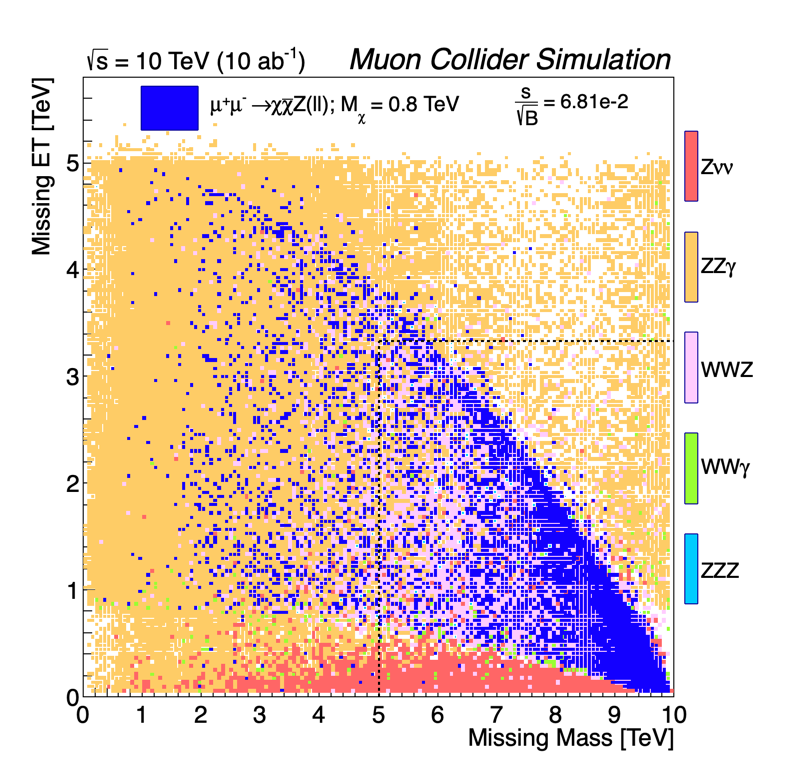}\\
  \includegraphics[width=2.5in]{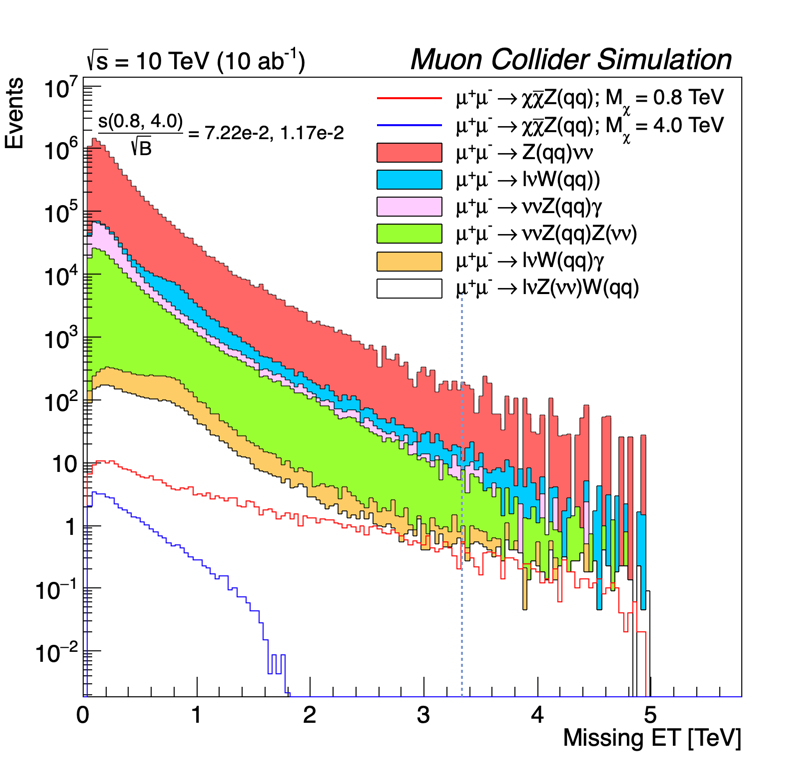}
  \includegraphics[width=2.5in]{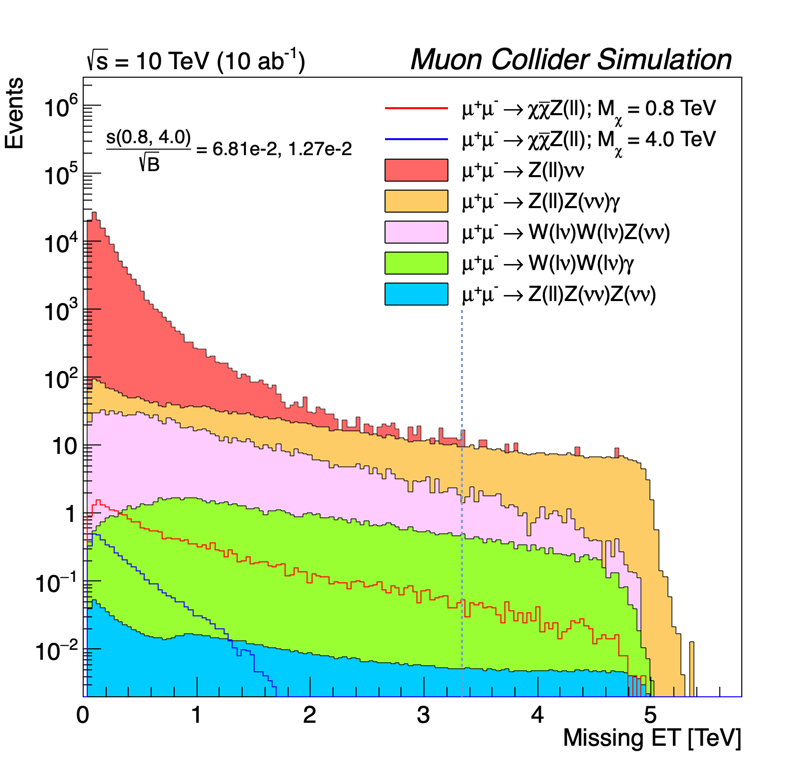} \\
    \includegraphics[width=2.5in]{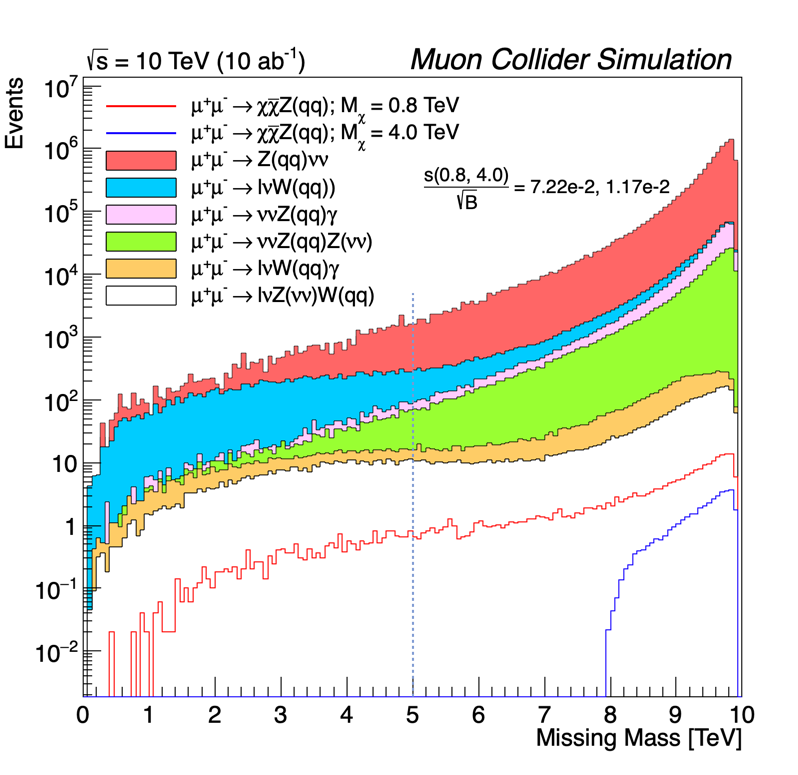}
    \includegraphics[width=2.5in]{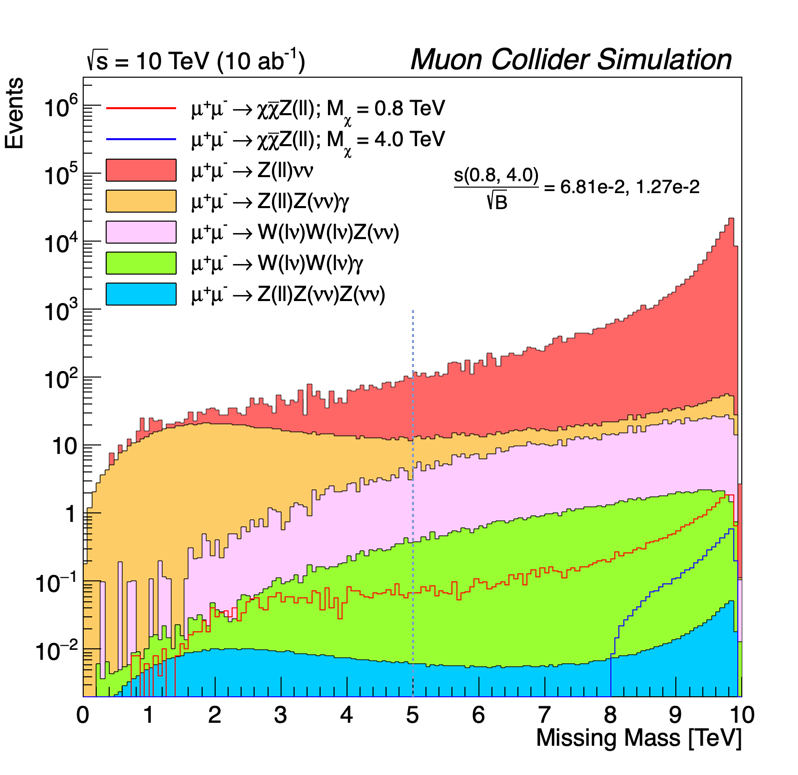} 
    \caption{The left side of the figure has plots for mono-Z($qq$) while right side has for mono-Z($\ell\ell$). Top panel: Scatter plot in missing transverse momentum versus missing mass plane for $m_\chi=0.8$ GeV, with proposed cuts indicated as the black rectangle. The simulated signal events in both scatter plot are shown in blue color and various simulated backgrounds are indicated in other colors. The signal significance (s/$\sqrt{B}$ is shown for an integrated luminosity of $10$ ab$^{-1}$ for the simulated $\sqrt{s}=10$ TeV data. The middle and bottom panel shows the MET and the missing mass projections. The signal and backgrounds in the MET and missing mass distributions are normalized to respective cross section and chosen value of luminosity}
    \label{fig:monoz-s10}
\end{figure}

\begin{figure}[htb]
    \centering
    \includegraphics[width=2.5in]{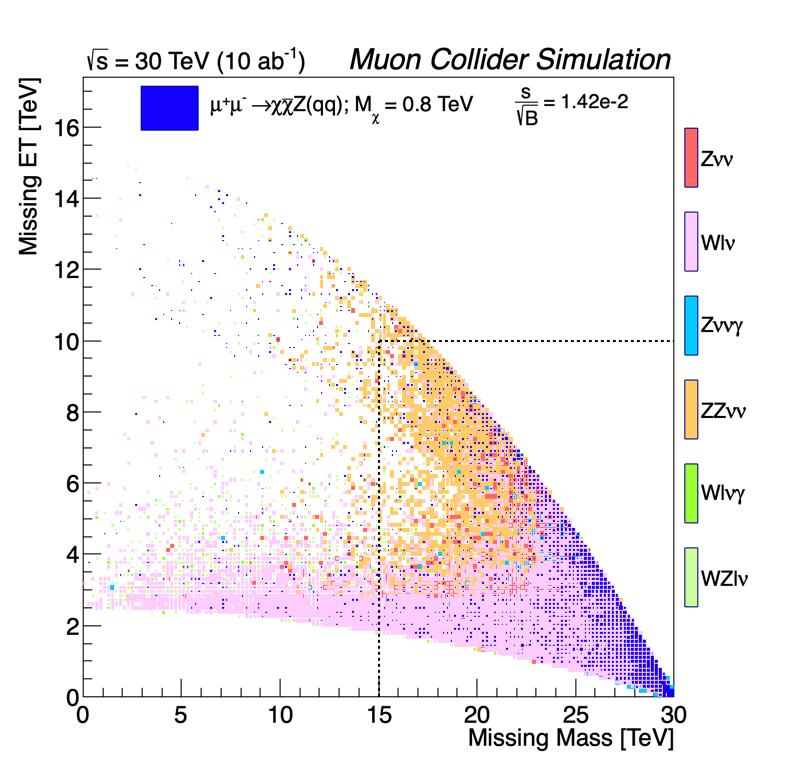}
    \includegraphics[width=2.5in]{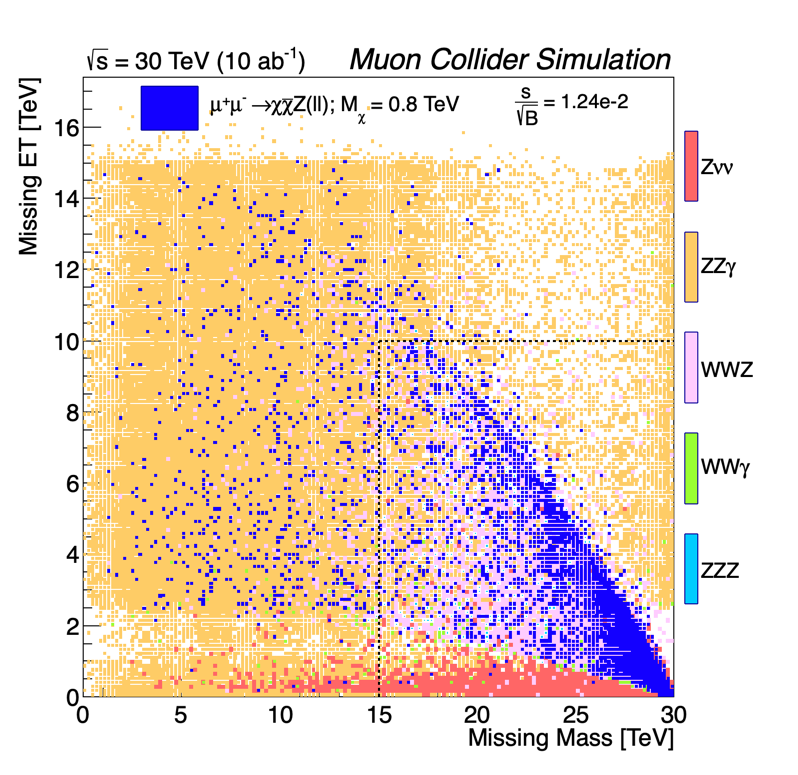}\\
  \includegraphics[width=2.5in]{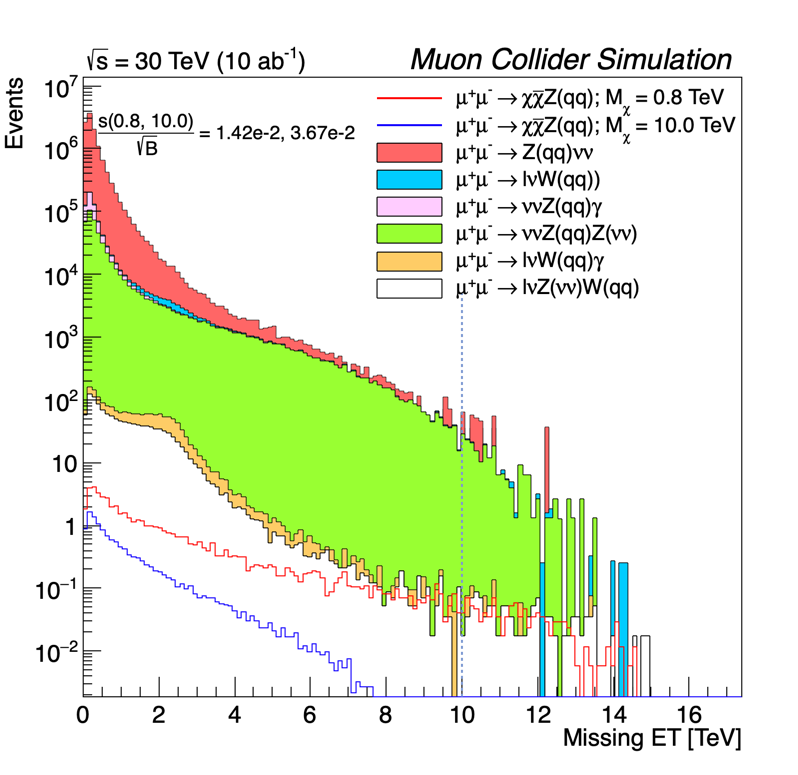}
  \includegraphics[width=2.5in]{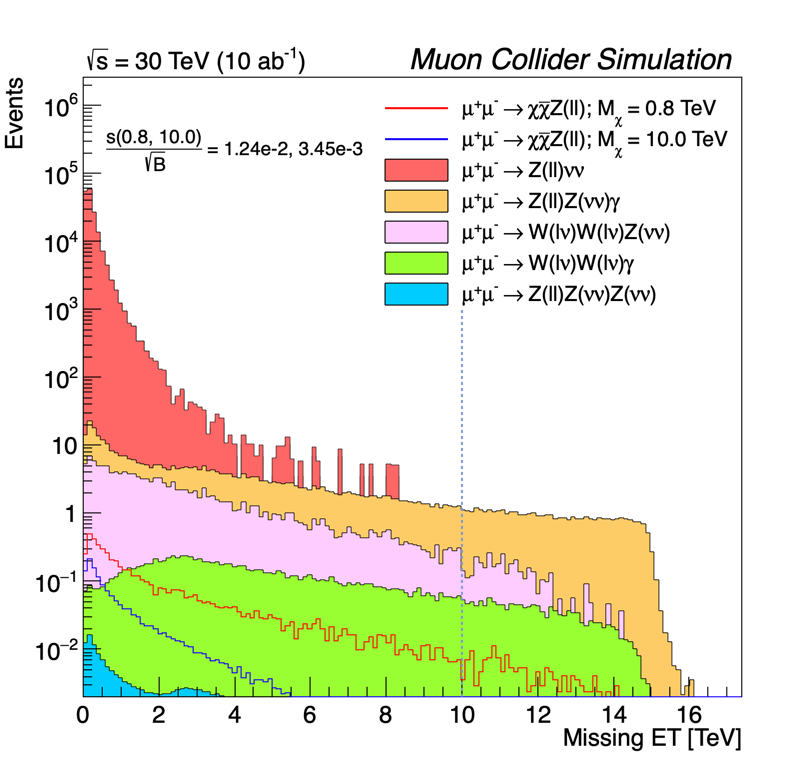} \\
    \includegraphics[width=2.5in]{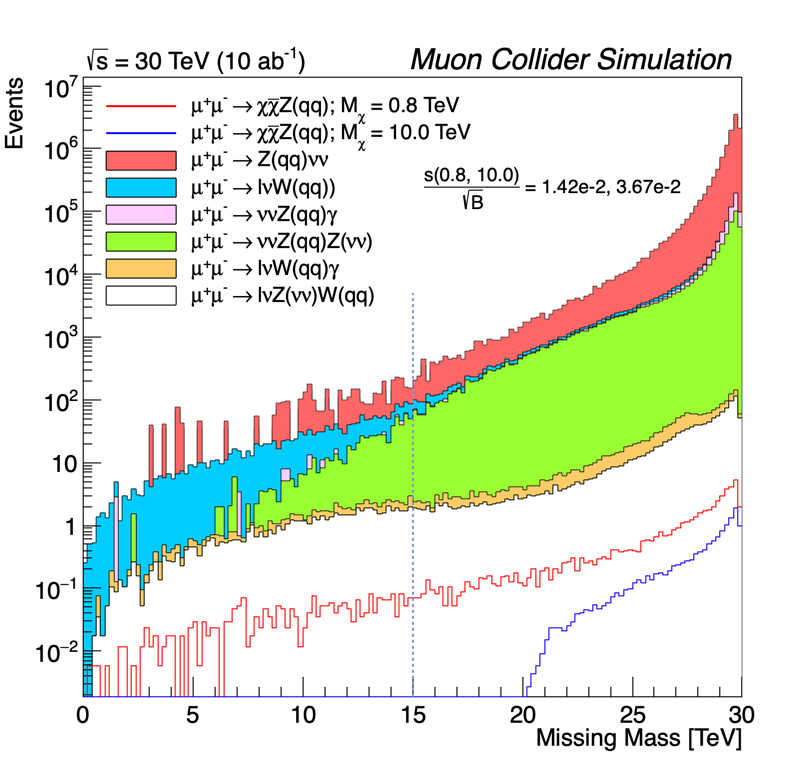}
     \includegraphics[width=2.5in]{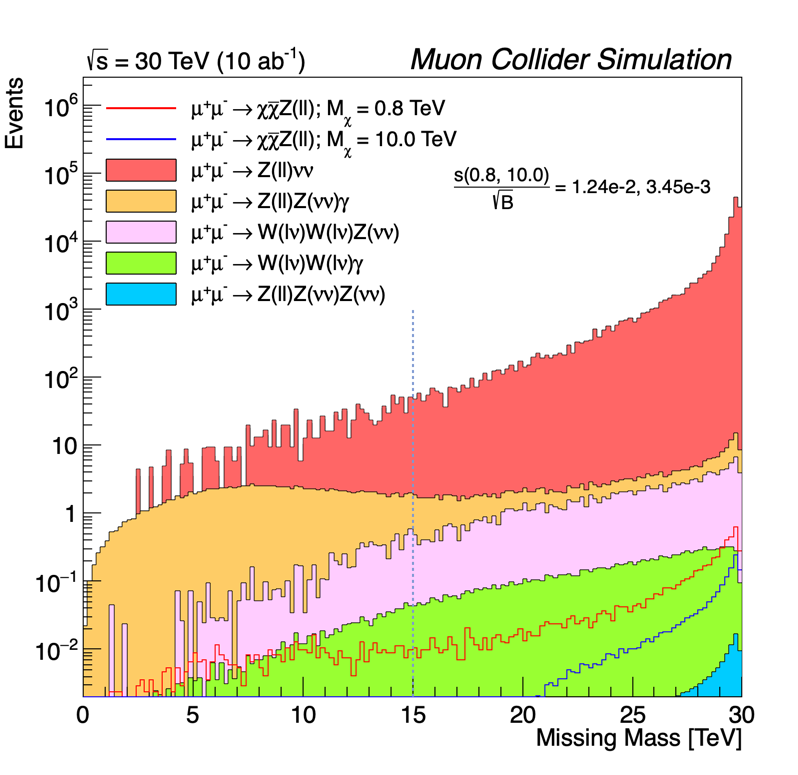}
    \caption{The left side of the figure has plots for mono-Z($qq$) while right side has for mono-Z($\ell\ell$). Top panel: Scatter plot in missing transverse momentum versus missing mass plane for $m_\chi=0.8$ GeV, with proposed cuts indicated as the black rectangle. The simulated signal events in both scatter plot are shown in blue color and various simulated backgrounds are indicated in other colors. The signal significance (s/$\sqrt{B}$ is shown for an integrated luminosity of $10$ ab$^{-1}$ for the simulated $\sqrt{s}=30$ TeV data. The middle and bottom panel shows the MET and the missing mass projections. The signal and backgrounds in the MET and missing mass distributions are normalized to respective cross section and chosen value of luminosity }
    \label{fig:monoz-s30}
\end{figure}

The Delphes generated signal events for each $E_{CM}$ and $m_\chi$ setting, and the associated backgrounds, are analyzed to select events with high MET in the final state and one Z boson that decays either hadronically or into a pair of leptons. In the case of an hadronically decaying Z-boson, events with at least two jets (or quarks) within the detector acceptance are considered, while for the leptonically decaying Z-boson, events with oppositely charged electrons or muons within the detector acceptance are considered. The invariant mass $M_{ll}$ of the leptons system is formed and further selection is applied to keep events with $60 < M_{ll} < 120$ GeV that are consistent with the Z boson mass. 
Using all the reconstructed tracks and neutral calorimeter objects in the event, i.e. those not matching with tracks, the event-level kinematic variables $\mathrm{MET}$ and $m_{missing}$ are reconstructed. 
The 2-dimensional distribution of the $\mathrm{MET}$ and $m_{missing}$ variables is shown in the top row of Figures~\ref{fig:monoz-s3}, \ref{fig:monoz-s6}, \ref{fig:monoz-s10}, and \ref{fig:monoz-s30} for different $E_{CM}$. In these distributions, the signal events are plotted in blue color. The dominant background due to Z-pair production, in which one of the Z bosons decays invisibly to neutrinos is shown in red. Sub-dominant backgrounds due to WWZ, ZZ$\gamma$, ZZZ and WW$\gamma$ production where some charged leptons could fall outside the detector acceptance are also shown. The signal and backgrounds events are scaled to the integrated luminosities of 1, 4, 10 and 10 ab$^{-1}$ for 3, 6, 10 and 30 TeV \mm~$E_{CM}$ using cross section values reported by the Madgraph5 generator for various production channels.
From these distributions, a selection in the 2-dimensional plane of $\mathrm{MET}$ and $m_{missing}$ is identified to increase the signal sensitivity to mono-Z events and are indicated by black lines on the figures.
The projections to the $\mathrm{MET}$ axis, after the $m_{missing}$ selection is shown in the middle panel of  Figures~\ref{fig:monoz-s3}, \ref{fig:monoz-s6}, \ref{fig:monoz-s10}, and \ref{fig:monoz-s30}. Similarly, projections to the $m_{missing}$ axis, after the $\mathrm{MET}$ selection are shown in the bottom panel. 
%The scatter plots of the $\mathrm{MET}$-$m_{missing}$ space are shown in top panel of the Figures~\ref{fig:monoz-s3}, \ref{fig:monoz-s6}, \ref{fig:monoz-s10}, and \ref{fig:monoz-s30}. 
%The signal events are plotted in green. The dominant background due to Z-pair production, in which one of the Z bosons decays invisibly to neutrinos is shown in red. The plot legend indicates the colors of the sub-dominant backgrounds due to $WWZ$, $ZZ\gamma$, $ZZZ$ and $WW\gamma$ production where some charged leptons could fall outside the detector acceptance. 
%The Monte Carlo statistics are large, but the backgrounds are scaled to the integrated luminosities of 1, 4, 10 and 10 ab$^-1$ for 3, 6, 10 and 30 TeV \mm~$E_{CM}$ using cross section values reported by the Madgraph5 generator for various production channels. 
%Cuts are selected in the 2D plane of $\mathrm{MET}$ and $m_{missing}$ for each case and are indicated by black lines on the figures. 
%The projections to the $\mathrm{MET}$ axis, after the $m_{missing}$-cut is shown in the bottom-left panes of the Figures~\ref{fig:monoz-s3}, \ref{fig:monoz-s6}, \ref{fig:monoz-s10}, and \ref{fig:monoz-s30}. Similarly,  projections to the$m_{missing}$ axis, after the $\mathrm{MET}$-cut are shown in the bottom-right pane of those plots. The numbers of signal and background events remaining after both the cuts are shown on the 2D plot. 

\begin{table}[hbt]
    \centering
    \renewcommand{\arraystretch}{1.2}
    \begin{tabular}{|l|c|c|c|c|c|c|c|c|}
        \hline
        \multirow{2}{2.5cm}{$m_\chi$ / $\sqrt{s}~(\int{d{\cal L}})$}& \multicolumn{2}{c}{3 TeV (1 ab$^{-1}$)} & \multicolumn{2}{|c}{6 TeV (4 ab$^{-1}$)} & \multicolumn{2}{|c}{10 TeV (10 ab$^{-1}$)} & \multicolumn{2}{|c|}{30 TeV (10 ab$^{-1}$)}\\
        \cline{2-9}
       % & $\sigma_{sig}$(fb) & $\sigma_{bkg}$(fb) & $\sigma_{sig}$(fb) &$\sigma_{bkg}$(fb) & $\sigma_{sig}$(fb) & $\sigma_{bkg}$(fb) & $\sigma_{sig}$(fb) & $\sigma_{bkg}$(fb)\\
       & $sig$ & $bkg$ & $sig$ & $bkg$ & $sig$ & $bkg$ & $sig$ & $bkg$\\
        \hline 
        0.5  &  95.2 & 7.0e5 &  n/a   &   n/a    &  n/a   &  n/a  &  n/a  &  n/a   \\
        0.6  &  82.5 & 7.0e5 &  n/a   &   n/a    &  n/a   &  n/a  &  n/a  &  n/a   \\
        0.7  &  71.7 & 7.0e5 &  n/a   &   n/a    &  n/a   &  n/a  &  n/a  &  n/a   \\
        0.8  &  59.1 & 7.0e5 &  162.8 &  3.64e6  &  230.1 & 1.01e7 &  48.5 & 1.17e7  \\
        1.0  &  34.8 & 7.0e5 &  143.6 &  3.64e6  &  211.8 & 1.01e7 &  46.1 & 1.17e7  \\
        2.0  &  n/a  &  n/a  &   58.9 &  3.64e6  &  144.8 & 1.01e7 &  39.1 & 1.17e7  \\
        2.5  &  n/a  &  n/a  &   21.9 &  3.64e6  &  115.9 & 1.01e7 &  36.3 & 1.17e7  \\
        4.0  &  n/a  &  n/a  &   n/a  &   n/a    &   37.4 & 1.01e7 &  30.1 & 1.17e7  \\
        10.0 &  n/a  &  n/a  &   n/a  &   n/a    &  n/a   &  n/a   &  12.6 & 1.17e7  \\ 
        \hline
    \end{tabular}
    \caption{The expected signal ($sig$) and background ($bkg$) events for center of mass energy of 3, 6, 10 and 30 TeV normalized to integrated luminosity of 1, 4, 10 and 10 ab$^{-1}$, respectively, for the mono-Z($qq$) process as a function of dark matter mass ($m_\chi$). The cross section for these processes are generated using MadGraph.}
    \label{tab:monoZqq-events}
\end{table}

\begin{table}[hbt]
    \centering
    \renewcommand{\arraystretch}{1.2}
    \begin{tabular}{|l|c|c|c|c|c|c|c|c|}
        \hline
        \multirow{2}{2.5cm}{$m_\chi$ / $\sqrt{s}~(\int{d{\cal L}})$}& \multicolumn{2}{c}{3 TeV (1 ab$^{-1}$)} & \multicolumn{2}{|c}{6 TeV (4 ab$^{-1}$)} & \multicolumn{2}{|c}{10 TeV (10 ab$^{-1}$)} & \multicolumn{2}{|c|}{30 TeV (10 ab$^{-1}$)}\\
        \cline{2-9}
    %    & $\sigma_{sig}$(fb) & $\sigma_{bkg}$(fb) & $\sigma_{sig}$(fb) &$\sigma_{bkg}$(fb) & $\sigma_{sig}$(fb) & $\sigma_{bkg}$(fb) & $\sigma_{sig}$(fb) & $\sigma_{bkg}$(fb)\\
        & $sig$ & $bkg$ & $sig$ & $bkg$ & $sig$ & $bkg$ & $sig$ & $bkg$\\
        \hline 
        0.5  &  11.9 & 9.63e3 &  n/a  &   n/a  &  n/a &  n/a   &  n/a  &  n/a   \\
        0.6  &  10.5 & 9.63e3 &  n/a  &   n/a  &  n/a &  n/a   &  n/a  &  n/a   \\
        0.7  &  9.18 & 9.63e3 &  n/a  &   n/a  &  n/a &  n/a   &  n/a  &  n/a   \\
        0.8  &  7.77 & 9.63e3 &  18.9 & 4.78e4 & 25.2 & 1.37e5 & 5.13  & 1.71e5 \\
        1.0  &  4.79 & 9.63e3 &  17.1 & 4.78e4 & 23.3 & 1.37e5 & 4.93  & 1.71e5 \\
        2.0  &  n/a  &  n/a   &  7.45 & 4.78e4 & 16.8 & 1.37e5 & 4.26  & 1.71e5 \\
        2.5  &  n/a  &  n/a   &  2.97 & 4.78e4 & 13.8 & 1.37e5 & 3.97  & 1.71e5 \\
        4.0  &  n/a  &  n/a   &  n/a  &   n/a  & 4.71 & 1.37e5 & 3.31  & 1.71e5 \\
        10.0 &  n/a  &  n/a   &  n/a  &   n/a  &  n/a &  n/a   & 1.42  & 1.71e5 \\
        \hline
    \end{tabular}
    \caption{The expected signal ($sig$) and background ($bkg$) events for center of mass energy of 3, 6, 10 and 30 TeV normalized to integrated luminosity of 1, 4, 10 and 10 ab$^{-1}$, respectively, for the mono-Z($\ell\ell$) process as a function of dark matter mass ($m_\chi$). The cross section for these processes are generated using MadGraph.}
    \label{tab:monoZll-events}
\end{table}

The sensitivity of this analysis to mono-Z events is quantified in terms of a figure of merit (FOM) defined as:

\begin{center}
    $FOM = \frac{\mathrm{s}}{\sqrt{\mathrm{b}}}$
\end{center}

where s is the numbers of expected signal events and b is the number of expected Standard Model background events after the final selection. The FOM obtained for different centers of mass energies of the collider and dark matter masses are listed in Table~\ref{tab:monoZqq-fom}. 
If only mono-Z processes are considered, a 10~TeV and 10-ab$^{-1}$ muon collider machine is needed to go significantly beyond the HL-LHC performance, with the current presented analysis. As future prospects, further improvements to the analysis could also be included to improve the discovery of DM through mono-Z events.

%The significance numbers are tabulated in \ref{tab:monoZqq-fom} for various accelerator settings and assumed dark matter masses. If only mono-Z channel is considered, a 10-TeV-10-ab$^{-1}$ machine is needed to go significantly beyond the HL-LHC performance, with the current non-optimal analysis. Further improvements and addition of hadronic Z-decays to the analysis could improve the reach. 

\begin{table}[htb]
    \centering
    \begin{tabular}{|l|c|c|c|c|}
        \hline
        &&&&\\
        $m_\chi$ / $\sqrt{s}~(\int{d{\cal L}})$ & 3 TeV (1 ab$^{-1}$) & 6 TeV (4 ab$^{-1}$) & 10 TeV (10 ab$^{-1}$) & 30 TeV (10 ab$^{-1}$)\\
        &&&&\\\hline
        0.5  &  1.14e-1 &   n/a    &   n/a    &   n/a     \\
        &&&&\\
        0.6  &  9.86e-2 &   n/a    &   n/a    &   n/a     \\
        &&&&\\
        0.7  &  8.57e-2 &   n/a    &   n/a    &   n/a     \\
        &&&&\\
        0.8  &  7.07e-2 &  8.53e-2 &  7.22e-2 &  1.42e-2  \\
        &&&&\\
        1.0  &  4.16e-2 &  7.52e-2 &  6.66e-2 &  1.34e-2  \\
        &&&&\\
        2.0  &   n/a    &  3.09e-2 &  4.55e-2 &  1.14e-2  \\
        &&&&\\
        2.5  &   n/a    &  1.15e-2 &  3.64e-2 &  1.06e-2  \\
        &&&&\\
        4.0  &   n/a    &   n/a    &  1.17e-2 &  8.82e-3  \\
        &&&&\\
        10.0 &   n/a    &   n/a    &   n/a    &  3.67e-3  \\
        \hline
    \end{tabular}
    \caption{Significance figure of merit for observing mono-Z($qq$) signals due to dark matter WIMP production above anticipated Standard Model backgrounds, as a function of dark matter mass ($m_\chi$), for muon colliders operating at various centers of mass and integrated luminosity.}
    \label{tab:monoZqq-fom}
\end{table}

\begin{table}[htb]
    \centering
    \begin{tabular}{|l|c|c|c|c|}
        \hline
        &&&&\\
        $m_\chi$ / $\sqrt{s}~(\int{d{\cal L}})$ & 3 TeV (1 ab$^{-1}$) & 6 TeV (4 ab$^{-1}$) & 10 TeV (10 ab$^{-1}$) & 30 TeV (10 ab$^{-1}$)\\
        &&&&\\\hline
        &&&&\\
        0.5  &  1.21e-1 &   n/a    &   n/a    &   n/a     \\
        &&&&\\
        0.6  &  1.07e-1 &   n/a    &   n/a    &   n/a     \\
        &&&&\\
        0.7  &  9.36e-2 &   n/a    &   n/a    &   n/a     \\
        &&&&\\
        0.8  &  7.91e-2 & 8.64e-2  & 6.81e-2  &   1.24e-2 \\
        &&&&\\
        1.0  &  4.89e-2 & 7.81e-2  & 6.31e-2  &   1.19e-2 \\
        &&&&\\
        2.0  &   n/a    & 3.41e-2  & 4.53e-2  &   1.03e-2 \\
        &&&&\\
        2.5  &   n/a    & 1.36e-2  & 3.73e-2  &   9.61e-3 \\
        &&&&\\
        4.0  &   n/a    &   n/a    & 1.27e-2  &   8.01e-3 \\
        &&&&\\
        10.0 &   n/a    &   n/a    &   n/a    &   3.45e-3 \\
        &&&&\\\hline
    \end{tabular}
    \caption{Significance figure of merit for observing mono-Z($\ell\ell$) signals due to dark matter WIMP production above anticipated Standard Model backgrounds, as a function of dark matter mass ($m_\chi$), for muon colliders operating at various centers of mass and integrated luminosity.}
    \label{tab:monoZll-fom}
\end{table}

\section{Summary}
Future high-energy muon-muon collider provide an interesting opportunity for dark matter discovery. Although strong astrophysical evidence indicates the existence of dark matter, there is no evidence yet for non-gravitational interactions between dark matter and standard model particles. These interactions, if present, can be studied at colliders and in particular at the future muon-muon collider. 

In this whitepaper, we presented studies for heavy weakly interacting massive particles part of a new electroweak multiplet with a high mass. In particular, we described prospects for WIMP discovery in mono-photon and mono-Z processes.

We have found that mono-photon signatures provide significant sensitivity to color-singlet-electroweak-doublet dark matter candidates, therefore providing a large impact in the search for the thermal dark matter. In order to target lower dark matter masses up to ~2 TeV, muon colliders should operate at low center of mass energies as preferred option, while the discover of more massive dark matter particles above 2 TeV would need higher energies and luminosities. The mono-Z channel, with the current presented analysis, would need a 10~TeV and 10-ab$^{-1}$ muon collider machine to go significantly beyond the HL-LHC performance. Despite being currently less sensitive with respect to the mono-photon signature, it could provide important complementary in the search for dark matter at muon colliders. In addition, possible future improvements in the analyses strategy could boost the sensitivity reach for this channel. %its sensitivity further optimizing the analyses strategy as well as including additional final states. 

\section{Acknowledgements}

The authors wish to thank Zhen Liu and Xing Wang for their help with the modeling of the Electroweak Multiplet Dark Matter model for this study.

This work is supported by the research grants from the US Department of Energy (Award number DE-SC0017647) and the University of Wisconsin.

\printbibliography

\end{document}